\documentclass[aps,prb,twocolumn,superscriptaddress,floatfix]
{revtex4-1}
\usepackage{graphicx}
\usepackage{amsmath,amssymb}
\begin{document}
\title{Impurity Green's function in the five frequency model of
diffusion in the FCC host: General case and the limit of strong
impurity-vacancy binding}
\author{V. I. Tokar}
\affiliation{Universit{\'e} de Strasbourg, CNRS, IPCMS, UMR 7504,
F-67000 Strasbourg, France}
\date{\today}
\begin{abstract}
The impurity Green's function exact to the first order in the vacancy
concentration has been calculated in the framework of the five frequency
model (5FM). The solution in terms of determinant ratios has been obtained
with the use of the Cramer's rule.  The determinant sizes varied from
54 for the most general case to three for the four frequency model.
Both analytical and numerical techniques has been used to analyze the
solution.  Special attention has been devoted to the case of strong
impurity-vacancy (I-v) binding in order to substantiate the picture of
diffusion via bound I-v pairs developed earlier in a phenomenological
approach.  Complete agreement with the phenomenological theory has been
established thus providing its rigorous justification.  The solution has
been also applied to the calculation of the diffusional broadening of
the M\"ossbauer resonance in Fe\underline{Al} system and good agreement
with available experimental data and calculations within the encounter
model has been found. The decay of the density of the nearest neighbor
I-v pairs has been discussed in detail and suggested to be used as an
additional constraint on the parameters of the 5FM. The possibility
of experimental observation of the decay with the use of the positron
annihilation technique has been pointed out.
\end{abstract}
\maketitle 
\section{Introduction}
In Ref.\ \onlinecite{I} (in the following referred to as I) the
impurity Green's function (GF) of the five-frequency model (5FM)
of the vacancy-mediated diffusion\cite{lidiard,LECLAIRE1978} was
calculated in the limit of a strong impurity-vacancy (I-v) attraction.
The solution was based on the phenomenological approaches of Refs.\
\onlinecite{cowern1990,In/cu(001)} where the bound I-v pairs were treated
as quasiparticles which diffusion was shown to be responsible for the
non-Gaussian diffusion profiles (NGDPs) and the non-Fickian diffusion
experimentally observed in the bulk diffusion in silicon\cite{cowern1990}
and in the Cu(001) surface layer.\cite{In-V-attraction} Because
of the instability of the bound pairs, the diffusion proceeds via
repeated I-v associations and subsequent decays with the encounters
separated by long periods of immobility because of the low vacancy
concentration that usually does not exceed 0.1 at.\~\% even at the
melting point.\cite{vacanciesDB2014,vacancies_S} After large number of
encounters the impurity density distribution acquires the conventional
Gaussian shape but at the early stage of diffusion it exhibits
the exponential-looking tails characteristic of a defect-mediated
diffusion.\cite{Brummelhuis1988,cowern1990,toroczkai1997,In-V-attraction}

In I the phenomenological approach was implemented within the 5FM with
the model parameters taken from the first-principles calculations of
Refs.\ \onlinecite{wu_high-throughput_2016,Wu2017} and the universal
NGDPs\cite{cowern1990,cowern1991} were simulated.  However, in Refs.\
\onlinecite{Brummelhuis1988,toroczkai1997,Pd/Cu,Co/cu(001)} it was
shown both theoretically and experimentally that for the existence of
exponential profiles the I-v attraction is not indispensable and may
occur even under the I-v repulsion.\cite{Co/cu(001),Pd/Cu} Furthermore,
the picture of bound I-v pairs diffusing at distances much larger than
the lattice constant does not accord well with the numerical simulations
of Ref.\ \onlinecite{voglAlFe} where despite rather strong I-v binding
(0.29~eV) the I-v encounter amounted on average to only two exchanges
between the impurity and the vacancy before their separation. This is
only 50\% more than the tracer-vacancy exchanges in the self-diffusion
with no I-v attraction. So the question arises about the uniqueness
of interpretation of NGDPs in terms of the pair diffusion and of
interrelation of this mechanism with the alternatives that do not require
I-v attraction, as well as of the ways of making distinction between them.

The aim of the present paper is to justify the pair diffusion
picture suggested in I within a rigorous master equation and Green's
function formalism.\cite{koiwa1983,cond-mat/0505019} Our approach
will be based on generalization of the technique developed in Refs.\
\onlinecite{tahir-kheli,tahir-kheli2} for the cases of self-diffusion
and of the diffusion in a two frequency model. The method of solution
that will be developed in the present paper can also be applied to
diffusion in any elemental crystalline hosts and to more complex models
with further than nearest neighbor (NN) vacancy jumps and with arbitrary
but finite I-v interaction range. Such extensions has been suggested in
literature on the basis of physical arguments and the first-principles
calculations.\cite{LECLAIRE1978,wu_high-throughput_2016,Bocquet,NNN_jumps}

The text of the paper is organized as follows. In the next section an
infinite system of rate equations for the impurity GF accurate to the
first order in the vacancy concentration will be derived;  in Sec.\
\ref{solution} it will be reduced to a finite system of 54 equations
amenable to solution by Cramer's rule. In Sec.\ \ref{mossbauer} the
solution will be used for illustrative calculation of the diffusional
broadening of the M\"ossbauer resonance\cite{voglAlFe,Fe57Cu}
and for discussion in this context of the encounter model
(EM).\cite{encounter_model0,bender,vogl_qens_1996,voglAlFe}   In Sec.\
\ref{001} the finite system obtained in Sec.\ \ref{solution} will be
farther reduces to the system of 13 equations which are sufficient
for the study of macroscopic diffusion. In Sec.\ \ref{D-r-Dm} the
diffusion constant and the parameters of the tightly bound I-v pairs
will be studied numerically to confirm the analytic expressions derived
in the phenomenological approach. In Sec.\ \ref{general_case} the
complications that arise in the case of I-v interaction of arbitrary
strength will be discussed and a simpler four frequency model (4FM)
that makes possible explicit analytic solution for impurity GF will be
introduced. Dissociation of NN pairs which may be present in the initial
state will be discussed in detail and a possibility of experimental
measurement of the evolution will be suggested.  It will be shown that
the data on the NN density evolution can provide additional information
about the values of the 5FM parameters. In final Sec.\ \ref{conclusion}
the main results will be briefly summarized.
\section{\label{exact}Rate equations for the impurity GF}
In the 5FM all interactions in the system comprising the host crystal,
the solute impurities, and the vacancies are reduced to interactions
within one I-v pair.\cite{lidiard,LECLAIRE1978,philibert,I} Hence, the
physics that the model can describe is that of diffusion in a dilute
alloy where each impurity can be treated independently, so theoretical
consideration can be restricted to only one of them. Because during one
I-v encounter the impurity is displaced only on a microscopic distance,
for macroscopic diffusion a large number of repeated encounters is
necessary.\cite{encounter_model0} Therefore, the ``minimal model'' that
will be studied below is a host crystal consisting of $N\to\infty$
sites containing one substitutional impurity and $c_vN$ vacancies
distributed within the crystal with concentration $c_v\ll1$. The large
number of vacancies makes the problem a truly many-body one but the small
vacancy concentration makes the solution of the first order in $c_v$
in most cases to be sufficient. Besides, the interactions of impurity
with two and more vacancies would be very different from the simple
superposition of independent I-v interactions and would require a large
number of additional parameters to be introduced, such as the values of
multi-vacancy interactions. This would strongly diminish the usefulness of
the model\cite{voglAlFe} so in the absence of adequate description of the
impurity interacting with more than one vacancy the physically meaningful
solution should not go beyond the pair I-v interactions. This reduces
the task to essentially a two-body problem that can be solved exactly.
Such a solution will be presented below and from a physical point of
view it may be considered as the best possible one because it cannot be
improved further without redefinition of the model.

Because 5FM is a stochastic model, it can be exactly described by
a suitable master equation (ME) (see Ref.\ \onlinecite{van_kampen}
and Appendix \ref{master_eq}). Its direct solution, however, usually
is efficient only in the case of a few-body system, like the bound
two-particle I-v pair treated in I with the use of ME. In the case of the
infinite number of vacancies a straightforward use of ME would require
dealing with the coordinates of all $c_vN=O(N)$ vacancies in the system
which is superfluous and practically unfeasible. An appropriate way of
proceeding is to average over positions of all vacancies that do not
participate in the I-v encounter. This can be achieved within the rate
equations (REs) approach which is, in fact, a method of solution of the
ME, as explained in Appendix \ref{master_eq}.

To derive the necessary REs let us define the impurity GF $G_{\bf l}(t)$
as the probability density of finding the impurity at time $t$ at the
lattice site described by vector ${\bf l}$ with the integer coordinates
$(l_X,l_Y,l_Z)$ if at time $t=0$ it was located at site ${\bf 0}=(0,0,0)$
\begin{equation}
	G_{\bf l}(t=0)=\delta_{\bf l0}.
	\label{Gini}
\end{equation}
This choice of the initial position was made for convenience because
GF for arbitrary initial position ${\bf l}^\prime$ can be trivially
obtained by the substitution ${\bf l}\to{\bf l-l}^\prime$.

To express GF as a statistical average let us introduce the occupation
number $\hat{\imath}_{\bf l}$ which is equal to unity when site ${\bf l}$
is occupied by the impurity and is zero otherwise.  In this notation
\begin{equation}
	G_{\bf l}(t)=\langle \hat{\imath}_{\bf l}\rangle
	\label{G(t)}
\end{equation}
where the angular brackets denote the average over the time-dependent
nonequilibrium statistical ensemble defined in Eq.\ (\ref{average}).
Thus, besides $c_vN$ vacancies the ensemble contains the inhomogeneous
impurity density distribution $G_{\bf l}(t)$. The evolution continues
until the impurity occupies all sites of the crystal with equal
probability $1/N$ in which case the density gradient vanishes and
the macroscopic diffusion ceases.

Because the I-v exchanges are restricted only to NN sites, the rate of change
of the impurity density on site ${\bf l}$ is determined solely by the
I-v configuration on this site and its NNs. Thus, the average impurity
density at site ${\bf l}$ may grow when the site is occupied by a vacancy
while the impurity occupies one of the NN sites, so that the impurity can
jump at the vacant site by the I-v exchange and the density may
diminish when site ${\bf l}$ is occupied by the impurity and there is a NN
vacancy for the impurity to leave the site. The exchanges occur with
frequency $w_2$ and their rates are proportional to the joint probability
density of simultaneous presence of the I-v pair at the NN sites as
\begin{eqnarray} 
\label{dG/dt}
\frac{d} {d t}G_{\bf l}(t) &=&\frac{d} {d t}\langle \hat{\imath}_{\bf l}\rangle
=w_{2}\sum_{\bf e}\left(\langle 
\hat{\imath}_{\bf l+{\bf e}}\hat{v}_{\bf l}\rangle
-\langle \hat{\imath}_{\bf l}\hat{v}_{\bf l+{\bf e}}\rangle
\right)\nonumber\\
&=&w_{2}\sum_{\bf e}[\rho_{\bf l+e}(t,{\bf -e})-\rho_{\bf l}(t,{\bf e})],
\end{eqnarray} 
where ${\bf e}$ is the set of 12 vectors connecting NN sites in the
FCC lattice and $\hat{v}_{\bf l}=0,1$ is the vacancy occupation number.
The joint pair probability density is defined as
\begin{equation}
\rho_{\bf l}(t,{\bf n})=\langle \hat{\imath}_{\bf l}\hat{v}_{\bf l+n}\rangle,
	\label{I-V}
\end{equation}
where ${\bf n}$ is the relative position of the vacancy with respect to
the impurity at site ${\bf l}$; the time dependence on the right hand side
(r.h.s.) is implicit in the nonequilibrium average Eq.\ (\ref{average}).

Thus, as could be expected on the basis of the general rate equation
Eq.\ (\ref{dA/dt}), the time derivative of the average impurity density
$\langle \hat{\imath}_{\bf l}\rangle$ in Eq.\ (\ref{dG/dt}) depends on
the average densities in Eq.\ (\ref{I-V}) so to solve Eq.\ (\ref{dG/dt})
further REs need be derived. Fortunately, it is easy to see that to the
first order in the vacancy concentration the chain of the REs terminates
already at the equation for the joint probability density $\rho$. Indeed,
because there is only one impurity in the system, higher order correlation
functions may contain only additional vacancy occupation numbers. But
then they will be of higher order in $c_v$ and thus can be omitted in
the $O(c_v)$ approximation.

The second (and the last in the approximation chosen) RE reads
\begin{eqnarray}\label{dpair/dt}
&&\frac{d} {d t}\rho_{\bf l}(t,{\bf n})
=w_{2}\sum_{\bf e}\delta_{\bf ne}[\rho_{\bf l+e}(t,{\bf -e})-\rho_{\bf l}(t,{\bf e})]\nonumber\\
&&+\sum_{\bf e}w_{\bf n+e,n}\rho_{\bf l}(t,{\bf n+e})
-\big(\sum_{\bf e}w_{\bf n,n+e}\big)\rho_{\bf l}(t,{\bf n}).
\end{eqnarray}
It is derived as follows. By the chain rule, the time derivative on the
r.h.s.\ of Eq.\ (\ref{I-V}) should consist of two contributions: one
corresponding to the change in the impurity density and the other one
in the vacancy density. Therefore, the first line on the r.h.s.\ of Eq.\
(\ref{dpair/dt}) should coincide with the r.h.s.\ of Eq.\ (\ref{dG/dt})
only corrected by the factor $\delta_{\bf ne}$ to account for the fact
that the impurity can move only when the only available vacancy at ${\bf
n}$ is at a NN site to the impurity. Similarly, the second line describes
the diffusion of a vacancy and formally has the same structure as the
equation for the free vacancy diffusion in Appendix \ref{free-vacancy}
but with an important difference that the jump frequencies are not all
equal to $w_0$ but depend on the vacancy position with respect to the
impurity and on the jump direction, as defined in the 5FM (see, e.\ g.,
Fig.\ 1 in I). In matrix notation it can be described with the use of
matrix $\tilde{W}$ having the same structure as matrix $\tilde{W}^0$ in
Eq.\ (\ref{W_vacancy}) but instead of $-12w_0$ at the matrix diagonal in
$\tilde{W}^0$ $\tilde{W}$ should contain the sum of all jump frequencies
from site $n$ to the 12 NN sites (the negative term on the last line of
Eq.\ (\ref{dpair/dt})) while the positive term should include frequencies
of all possible vacancy jumps from the sites that are NN to $n$ to this
site (in $\tilde{W}^0$ they are all equal to $w_0$).

As is seen, the equation for the rate of change of the pair density
Eq.\ (\ref{dpair/dt}) in $O(c_v)$ approximation depends only on the
pair density itself. So the density evolution can be found from this
equation alone, provided the initial condition is known. With the
initial location of the impurity at the coordinates origin, it remains
to chose the initial distribution of the vacancies. In principle, in
an out-of-equilibrium system the distribution can be arbitrary. But
in order to describe the evolution towards thermal equilibrium one
would need to know in detail the vacancy kinetics. For example, if the
vacancies are in excess in comparison with the equilibrium concentration
than in order to describe such phenomena as the creation of divacancies,
vacancy pores, or dislocation loops one would need to know intervacancy
interactions. Besides, both under the vacancy excess or deficit their
sources or sinks must be introduced into the model in order to correctly
describe the kinetics leading to the equilibrium. Such terms, however,
are absent in the 5FM, so the model implicitly presumes the vacancy
concentration to be a constant parameter independent of the kinetics. This
is possible only at thermal equilibrium. This reasoning, however, is valid
only for the global vacancy concentration while a local distribution in
the vicinity of the impurity still can be arbitrary. The local density
perturbation concerns only $O(1)$ number of vacancies which cannot
influence global concentration $c_v$ in the thermodynamic limit. But to
farther simplify the problem, in the present paper we will restrict our
consideration to the simplest symmetric distribution.  Namely, we assume
that at $t=0$ the vacancies are distributed homogeneously everywhere
in the crystal with the equilibrium density $c_v$ except at the site
occupied by the impurity where the density should be zero by the vacancy
definition and at the sites NN to the impurity where the concentration
$c_{NN}$ is assumed to be different from $c_v$ because, e.\ g., of the
I-v interaction.  These conditions are satisfied by the expression
\begin{equation}
\rho_{\bf l}(t=0,{\bf n})=\delta_{\bf l0}\big[c_v-c_v\delta_{\bf n0}
+(c_{NN}-c_v)\sum_{\bf e} \delta_{\bf ne}\big],
	\label{ini}
\end{equation}
where the first factor on the r.h.s.\ accounts for the initial
impurity position.

The initial value problem for a linear equation with constant coefficients
can be conveniently solved by means of the combined integral Laplace and
Fourier transform (the LF-transform) that for an arbitrary function $\phi$
can be defined as
\begin{equation}
	LF[\phi_{\bf l}(t)]=\phi({\bf K},z)=\int_0^\infty dt\,e^{-zt}
	\sum_{\bf l}e^{-ia{\bf K\cdot l}/2}\phi_{\bf l}(t).
	\label{LFdef}
\end{equation}
Under the transform Eq.\ (\ref{dpair/dt}) becomes
\begin{eqnarray}
	\label{eq_transformed}
&&z\rho({\bf K},z,{\bf n})=c_v-c_v\delta_{\bf n0}+(c_{NN}-c_v)
\sum_{\bf e} \delta_{\bf ne}\nonumber\\
&&+w_2\sum_{\bf e} \delta_{\bf ne}\big[e^{ia{\bf K\cdot e}/2}
\rho({\bf K},z,{\bf-n}) -\rho({\bf K},z,{\bf n})\big]\\
&&+\sum_{\bf e}w_{\bf n+e,n}\rho({\bf K},z,{\bf n+e})
-\bigg(\sum_{\bf e}w_{\bf n,n+e}\bigg)\rho({\bf K},z,{\bf n}),\nonumber
\end{eqnarray}
where
\begin{equation}
	\rho({\bf K},z,{\bf n})	= LF[\rho_{\bf l}(t,{\bf n})].
	\label{directLF}
\end{equation}
The first line on the r.h.s.\ in Eq.\ (\ref{eq_transformed}) is the
Fourier transformed initial value of the pair density Eq.\ (\ref{ini}).
The remaining two lines originate from the first and the second lines on
the r.h.s.\ of Eq.\ (\ref{dpair/dt}). Thus, for each ${\bf K}$ we obtain
an infinite inhomogeneous system of equations for the $N$-dimensional
vector $\vec{\rho}_{\bf K}$ with components
\begin{equation}
	\vec{\rho}_{\bf K}|_{\bf n}
={\rho}({\bf K},z,{\bf n})  
	\label{vec_rho}
\end{equation}
which in the thermodynamic limit becomes infinite-dimensional.  In the
next section it will be shown that if the jump rates $w_{\bf n,m}$ in
Eq.\ (\ref{eq_transformed}) differ from the bulk rate $w_0$ only within
a finite region around the impurity, the infinite set of equations can
be reduced to a finite linear system.

To conclude this section let us derive an expression for the
impurity GF in terms of the solution of Eq.\ (\ref{eq_transformed}).
The LF-transformed Eq.\ (\ref{dG/dt}) reads
\begin{equation}
	zG({\bf K},z)=1-w_2\sum_{\bf e}(1-e^{-ia{\bf K\cdot e}/2})
	\rho({\bf K},z,{\bf e}),
	\label{zG}
\end{equation}
where use has been made of the fact that summations over ${\bf e}$
and $-{\bf e}$ are equivalent due to the lattice symmetry. Casting Eq.\
(\ref{zG}) in the form 
\begin{equation}
	G({\bf K},z)=\frac{1}{z}+\frac{1}{z^2}\Sigma({\bf K},z),
	\label{zG2}
\end{equation}
where 
\begin{equation}
	\Sigma({\bf K},z)=-w_2\sum_{\bf e}(1-e^{-ia{\bf K\cdot e}/2})
z\rho({\bf K},z,{\bf e}),
	\label{SzK}
\end{equation}
is the diffusion kernel, one can recognize in Eq.\ (\ref{zG2}) the first two
terms of the expansion of of the GF in the Dyson form
\begin{equation}
	G({\bf K},z)=\frac{1}{z-\Sigma({\bf K},z)},
	\label{dyson}
\end{equation}
as discussed in Sec.\ IV in I.  In the theory of vacancy-mediated
diffusion Eq.\ (\ref{dyson}) can also be obtained in the rate equation
approach.\cite{bender}  In Appendix \ref{dyson_eq} we show that
it can be rigorously derived in the framework of the Mori-Zwanzig
memory-function formalism.  But the equations for the memory function
(or the self-energy\cite{forster}) that can be obtained from Eq.\
(\ref{dyson4}) become useful mainly in high order expansions where they
significantly reduce the number of contributions by restricting them only
to irreducible ones. The lowest order term that we are interested
in is easily recoverable from the much simpler REs for GF.
\section{\label{solution}Solution of the rate equations by the Cramer's
rule}
The method of solution presented below is a generalization
to the case of the 5FM of the method suggested in Refs.\
\onlinecite{tahir-kheli,tahir-kheli2} for the cases of self-diffusion and
for the impurity diffusion in a two-frequency model. As will be seen, the
method can be applied to any model with I-v interactions of finite range.
Its essence is the reduction of the infinite (in the thermodynamic limit)
set of equations Eq.\ (\ref{eq_transformed}) to a finite linear system.

To begin with, we note that one component of $\vec{\rho}_{\bf K}$
can be found immediately. Examination of Eq.\ (\ref{eq_transformed}) shows
that $\rho({\bf K},z,{\bf n=0})=0$ satisfies the equation independently
of the values of other components of $\vec{\rho}_{\bf K}$. Indeed, only
the terms on the third line on the r.h.s.\ of Eq.\ (\ref{eq_transformed})
could potentially cause problems. But because simultaneous presence of
the vacancy and the impurity at the same site are forbidden, the jump
rates to- and from the impurity site must be set equal to zero
\begin{equation}
	w_5=w_{\bf e,0}=w_{0,\bf e}=0.	
	\label{w_5}
\end{equation} 
Here the fictitious sixth frequency $w_5=0$ was introduced
in order to formally treat the jumps between all sites in the
lattice on the same grounds. Thus, $\rho({\bf K},z,{\bf n=0})=0$
satisfies Eq.\ (\ref{eq_transformed}) and later will be excluded from
consideration. 

Our next step is to subtract from both sides of Eq.\
(\ref{eq_transformed}) vector $\tilde{W}^0\vec{\rho}_{\bf K}$ with
the components
\begin{equation}
	\left.\tilde{W}^0\vec{\rho}_{\bf K}\right|_{\bf n}=	
	\sum_{\bf m}\tilde{W}_{\bf nm}^0\rho({\bf K},z,{\bf m}),
	\label{Wpsi}
\end{equation}
where $\tilde{W}^0$ is the matrix from Eq.\ (\ref{W_vacancy}) that
describes the vacancy diffusion in the pure host crystal.  The last
line in Eq.\ (\ref{eq_transformed}) has similar formal structure and
can be written down as $\tilde{W}\vec{\rho}_{\bf K}$ so that after the
subtraction the terms on the line become
\begin{equation}
(\tilde{W}-\tilde{W}^0)\vec{\rho}_{\bf K}.
	\label{tilded}
\end{equation}
Because beyond the forth coordination shell (CS) the rates in $\tilde{W}$
are equal to those of $\tilde{W}^0$ (see the discussion following
Eq.\ (\ref{dpair/dt})), the matrix difference in Eq.\ (\ref{tilded})
has nonzero matrix elements only for ${\bf m,n}$ in a vicinity of
the impurity.  This means that Eq.\ (\ref{tilded}) contains only
a finite number of components of $\vec{\rho}_{\bf K}$. Taking into
account that ${\rho}_{\bf K}({\bf n})$ on the second line in Eq.\
(\ref{eq_transformed}) are restricted to the first CS, the number of
the vector components on the r.h.s.\ of Eq.\ (\ref{eq_transformed})
is also finite.

\begin{figure}
\begin{center}
\includegraphics[viewport = 0 10 400 400, scale = 0.4]{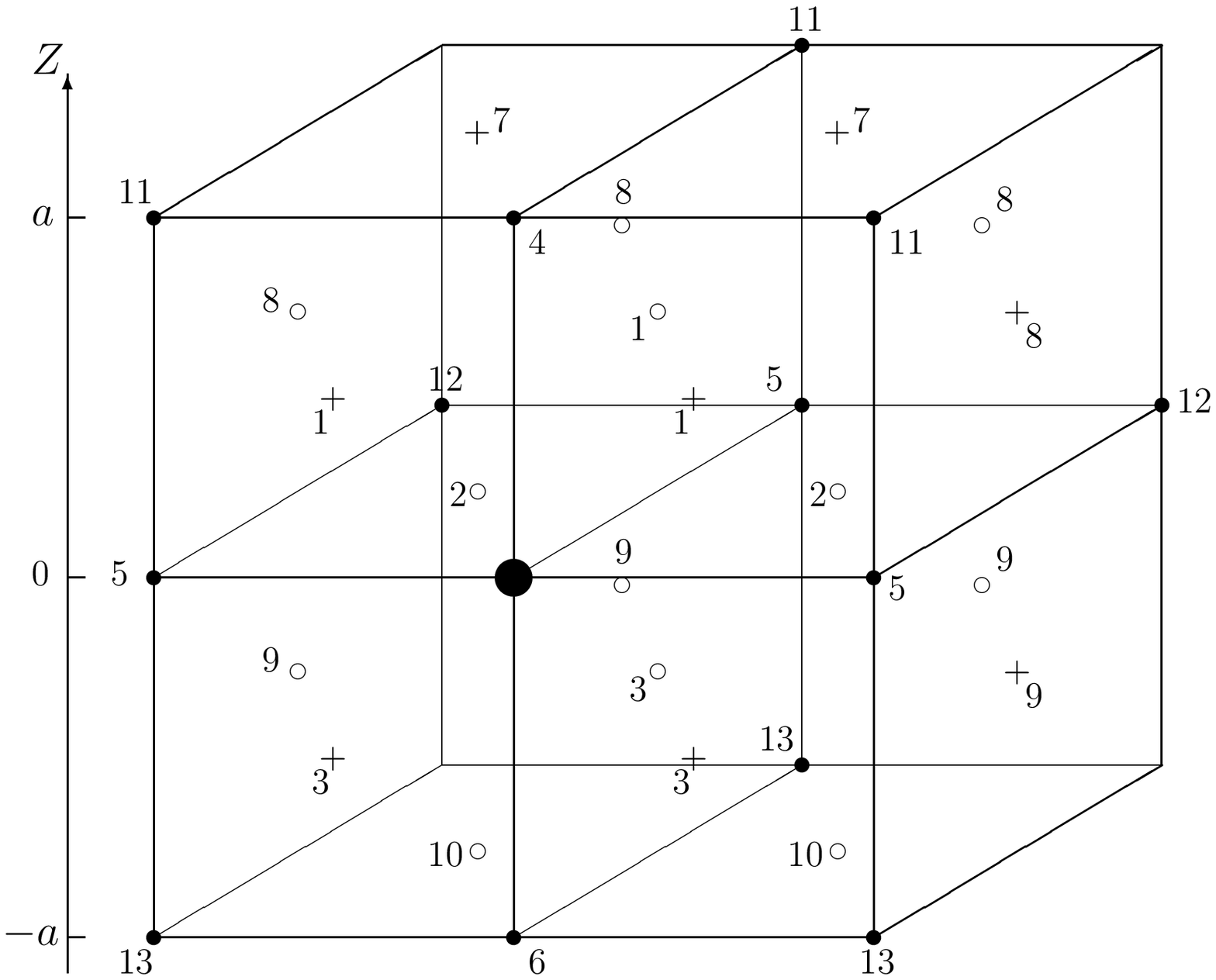}
\end{center}
\caption{\label{fig1-2}Axisymmetric diffusion along $Z$ direction of the
FCC lattice as described by the 5FM. The lattice sites in the vicinity
of the impurity (black circle) belong to 13 equivalence classes marked
as: Black points---the classified sites in the vertexes of large cubes;
crosses---sites at the centers of the external faces of the cubes in the 
drawing, open circles---on the internal faces.}
\end{figure}

As can be seen from the definition of the 5FM and from Fig.\
\ref{fig1-2}, there are 54 sites in four CSs plus one site at ${\bf
n=0}$ (55 in total\cite{koiwa1983}).  In matrix notation the equation
can now be written as
\begin{equation}
(z-\tilde{W}^0)\vec{\rho}_{\bf K}=\vec{c}_{0}+V\vec{\rho}_{\bf K},
	\label{rhs}
\end{equation}
where in the $55\times55$ matrix $V$ we gathered all matrix elements
of Eq.\ (\ref{tilded}) and also the terms on the second line of Eq.\
(\ref{eq_transformed}). The inhomogeneous term is the vector composed
of the terms on the first line of Eq.\ (\ref{eq_transformed})
\begin{equation}
	\vec{c}_{0}|_{\bf n}=c_v-c_v\delta_{\bf n0}+(c_{NN}-c_v)
\sum_{\bf e} \delta_{\bf ne}.
	\label{c_0}
\end{equation}

However, on the l.h.s.\ of Eq.\ (\ref{rhs}) all components of the vector
$\vec{\rho}_{\bf K}$ are still present after the subtraction:
\begin{equation}
	(z-\tilde{W}^0)\vec{\rho}_{\bf K}.
	\label{lhs}
\end{equation}
The components are mutually coupled because of the structure of matrix
$\tilde{W}^0$ in Eq.\ (\ref{W_vacancy}) which always connects the vector
components in a given CS to higher CSs. To overcome this difficulty,
let us multiply both sides of the equation by the matrix
\begin{equation}
	\tilde{P}=(z-\tilde{W}^0)^{-1}=\left[P_{{\bf n-n}^\prime}\right]
	\label{tildeP}
\end{equation}
with the matrix elements given by Eq.\ (\ref{P}). Under the
multiplication the number of unknown components of $\vec{\rho}_{\bf K}$ 
on the r.h.s.\ will not change but on the l.h.s.\ the components now
decouple from each other
\begin{equation}
	\vec{\rho}_{\bf K}=\tilde{P}\vec{c}_{0}+\tilde{P}V\vec{\rho}_{\bf K}.
	\label{timesP}
\end{equation}
So by retaining only those equations that contain on the l.h.s.\ the same
vector components that are present on the r.h.s.\ one arrives at a finite
system of linear equations which in the conventional form reads
\begin{equation}
	(I-\tilde{P}V)\vec{\rho}_{\bf K}=\tilde{P}\vec{c}_{0},
	\label{canonical}
\end{equation}
where $I$ is the unit matrix.  The size of the system can be restricted to
54 equations because the component $\rho_{\bf K}({\bf n=0})=0$ is already
known. As is easy to see, to obtain the $54\times54$ matrix $\tilde{P}V$
for the reduced equation it is sufficient to multiply $54\times55$
matrix $P$ by $55\times54$ matrix $V$ by omitting the row and the
column corresponding to ${\bf n=0}$ in the initial $55\times55$ matrices
$P$ and $V$, respectively.  The free term on the r.h.s.\ can be found
explicitly with the use of Eqs.\ (\ref{eq_transformed}), (\ref{P}), and 
(\ref{relations}) as
\begin{equation}
	(\tilde{P}\vec{c}_{0})_{\bf n}=\frac{c_v}{z}+[(c_{NN}-c_v)
(12+\frac{z}{w_0})-c_v]P_{\bf n}.
	\label{Prho}
\end{equation}
With the matrix on the l.h.s.\ and the inhomogeneous term given by
Eq.\ (\ref{Prho}) being known explicitly, Eq.\ (\ref{canonical}) can
now be solved by the standard means of linear algebra.  From Eqs.\
(\ref{SzK}) and (\ref{dyson}) it is seen that for the calculation of
the impurity GF for general Fourier momentum ${\bf K}$ only twelve NN
components of $\vec{\rho}_{\bf K}$ are needed. But because FCC lattice
is centrosymmetric this number can be reduced to six, so the use of
the Cramer's rule seems to be computationally simpler than the full
matrix inversion.

The GF obtained from the solution of the 54 equations
Eq.\ (\ref{canonical}) can be used to describe
situations where large values of momentum ${\bf K}$ are of
interest,\cite{vogl_qens_1996,leitner_atomic_2009,voglAlFe,Fe57Cu}
which will be illustrated in the next section.
\section{\label{mossbauer}Diffusional broadening of the M\"ossbauer line}
An important advantage of the rigorous solution for the impurity GF
is that it is not restricted to small values of the Fourier momentum
as the phenomenological GF but is valid for arbitrary ${\bf K}$.
This makes possible theoretical description of such techniques as the
quasielastic neutron scattering, the coherent X-rays and the M\"ossbauer
spectroscopies.\cite{vogl_qens_1996,leitner_atomic_2009,voglAlFe,Fe57Cu}

As an illustrative example let us consider the diffusional broadening
of the M\"ossbauer line in the Fe\underline{Al} system studied in Ref.\
\onlinecite{voglAlFe}. The system is interesting from the point of
view of the pair diffusion because of the rather strong I-v attraction
of 0.29~eV as estimated from the experimental data.\cite{voglAlFe}
However, the diffusion profiles simulated on the basis of the extracted
5FM parameters were found to be roughly Gaussian (see Fig.\ 3 in Ref.\
\onlinecite{voglAlFe}). Furthermore, some other conclusions drawn by
the authors do not agree with the results obtained in I, so below we
will attempt to clarify these issues.

Diffusional behavior contributes to the M\"ossbauer broadening
through the van Hove correlation function which at small impurity
concentrations coincides with the single impurity GF calculated in
Sec.\ \ref{solution}. In the M\"ossbauer studies it was found that the
line broadening can be adequately described in the framework of the
EM\cite{encounter_model0,bender,vogl_qens_1996,voglAlFe} so we first
establish connection of our approach with the EM. To this end we note that
an important parameter of the EM is the average number of steps $z_{enc}$
that the impurity makes during one I-v encounter.  Physically $z_{enc}$
is akin to the mean diffusion distance $\lambda$ of the phenomenological
theory and establishing a formal relationship between the parameters
should facilitate comparison between two approaches.

Formally the diffusional contribution to the M\"ossbauer line
broadening is given by the real part of the impurity GF at
$z=\gamma/2+i\omega$, where $\gamma$ is the natural width of the
M\"ossbauer line and $\hbar\omega$ is the energy transferred to
the system by the gamma ray, so our solution for the GF should be
sufficient for the task.\cite{bender,voglAlFe} However, in Refs.\
\onlinecite{bender,voglAlFe} it was pointed out that the solutions of
the kind of our Eq.\ (\ref{dyson}) cannot be directly compared with
experimental M\"ossbauer spectra because the measured quantity is
the width of the line which can be accurately fitted by the Lorentzian
distribution.  But the line shape in Eq.\ (\ref{dyson}) is not Lorentzian
because of the $z$-dependent diffusion kernel $\Sigma$. The difficulty
is overcome by the EM that describes the diffusion as a sequence of
repeated I-v encounters.\cite{encounter_model0,bender} In the course
of the encounter a vacancy is always present in the vicinity of the
impurity so all impurity jumps occur within a short time interval $\Delta
t_{enc}=O(1)$ which is small in comparison with the time interval between
the encounters that is equal to the association time
\begin{equation}
	t_a=g^{-1}=O(c_v^{-1})\gg \Delta t_{enc}.
	\label{t_enc}
\end{equation}
So to a good approximation $\Delta t_{enc}$ can be neglected on the scale
of $t_a$ and the impurity transfer during the encounter from the initial
to the final position may be considered as instantaneous. This picture
can be translated into the following approximation to GF. The diffusion
equation satisfied by GF in Eqs.\ (\ref{dyson}) and (\ref{dyson3})
under the inverse LF-transform would read
\begin{equation}
	\partial G_{\bf l}(t)/\partial t=\int_0^t
         dt^\prime\sum_{\bf l^\prime}
	\Sigma_{{\bf l}^\prime}(t^\prime)G_{\bf l-l^\prime}(t-t^\prime).
	\label{DeqS}
\end{equation}
Because $\Sigma$ varies on the time scale $O(\Delta t_{enc})$ which is
much shorter than $t$, the self-energy will differ from zero only in a
narrow region of its argument where $t^\prime\ll t$ so in the argument of
$G$ in Eq.\ (\ref{DeqS}) $t^\prime$ can be neglected: $t-t^\prime\approx
t$. Now the integration over $t^\prime$ reduces to multiplication of
GF at time $t$ by $\int_0^t\Sigma(t^\prime)$ where the upper limit can
be safely extended to infinity.\cite{bender} But such integration is
equivalent to $z=0$ component of the Laplace transform of $\Sigma$,
so the LF-transformed GF in this approximation will be
\begin{equation}
	G^{enc}({\bf K},z)=\frac{1}{z-\Sigma({\bf K},0)}  \label{Genc}
\end{equation} 
(the correction due to the natural line width $\gamma$ was found to be
negligible in the experimental conditions of Ref.\ \onlinecite{voglAlFe}).
In this approximation the real part of GF becomes a Lorentzian of
width\cite{bender}
\begin{equation}
	\Delta\Gamma=-2\Sigma({\bf K},0),
 \label{DGamma}
\end{equation} 
and is a real non-negative function of ${\bf K}$.

\begin{figure}
\begin{center}
\includegraphics[viewport = 170 10 220 240, scale = 0.6]{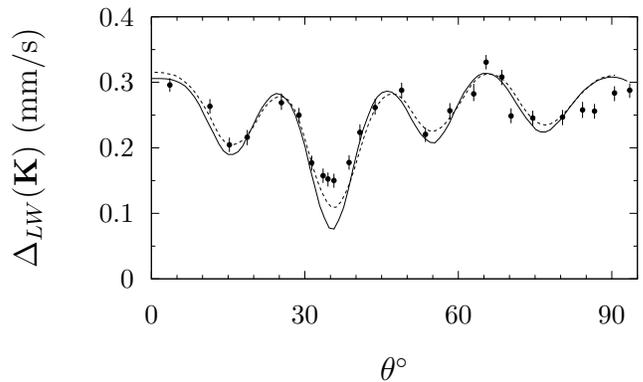}
\end{center}
\caption{\label{fig2-2}Symbols and the dashed line: experimental data
and the calculated anisotropy of the M\"ossbauer resonance from Ref.\
\onlinecite{voglAlFe}; solid line: Eq.\ (\ref{Delta_LW}) with the same
parameters as in Ref.\ \onlinecite{voglAlFe} except the value of $w_2$
being reduced on 4\%, as explained in the text.}
\end{figure}

The quantity\cite{voglAlFe}
\begin{equation}
\Delta_{LW}({\bf K})=\Delta\Gamma/|{\bf K}|=-2\Sigma({\bf K},0)/|{\bf K}|
	\label{Delta_LW}
\end{equation}
with ${\bf K}$ corresponding to the M\"ossbauer gamma rays and with
$\Sigma$ from the Cramer's rule solution of Sec.\ \ref{solution}
has been calculated and smeared with the experimental resolution. In
Fig.\ \ref{fig2-2} the curve thus obtained is compared with
the experimental data and the Monte Carlo (MC) simulations of Ref.\
\onlinecite{voglAlFe}. The same parameters of the 5FM were used in the
calculations with a minor correction for $w_2$ which was reduced by 4\%
to account for the fact that in the MC simulations the correlation factor
was estimated as $f=0.92$ while Eq.\ (\ref{f}) gives $f=0.96$. So $w_2$
was modified for the calculated diffusion constant agreed with the fit
to experimental data.\cite{voglAlFe}

In the EM the momentum-dependent line broadening was found to 
be (see Eqs.\ (6) and (7) in Ref.\ \onlinecite{encounter_model0})
\begin{equation}
	\Delta\Gamma=\frac{2}{\tau_{enc}}\left(1-W^{enc}({\bf K})\right),
	\label{Gamma-W}
\end{equation}
where $W^{enc}$ is the Fourier transform of the distribution of the
impurity density over the lattice sites after the I-v encounter
with $\tau_{enc}$ being the average time between the encounters.
In the phenomenological approach similar expression can be obtained
by substituting Eq.\ (\ref{sigma_K}) from Appendix \ref{parameters}
into Eq.\ (\ref{DGamma}):
\begin{equation}
	\Delta\Gamma=2g\left(1-\frac{1}{1+(\lambda{\bf K})^2}\right).
	\label{Gamma-g}
\end{equation}
In the phenomenological approach $\tau_{enc}$ is equal to the association
time $t_a=1/g$ so comparing Eqs.\ (\ref{Gamma-g}) and (\ref{Gamma-W})
one may conclude that
\begin{equation}
	W^{enc}({\bf K})\simeq 1/[1+(\lambda{\bf K})^2].
	\label{Wenc-approx}
\end{equation}
The equality here may be achieved only for asymptotically small $|{\bf
K}|\to0$ in the region of validity of the phenomenological theory.
In the M\"ossbauer experiment, on the other hand, $|{\bf K}|$ is finite
and comparatively large so $W^{enc}$ in Eq.\ (\ref{Wenc-approx}) cannot
be found in the small-${\bf K}$ limit.  However, below we will see that
the number of impurity jumps $z_{enc}$ depends only on the small momenta,
Eq.\ (\ref{Wenc-approx}) should be sufficient for establishing an exact
relation between $z_{enc}$ and $\lambda$.

The relationship can be found with the use of the expression
\begin{equation}
	z_{enc}=\frac{1}{f}\sum_{{\bf l}}W^{enc}_{\bf l} \frac{{\bf
	l}^2}{{\bf e}^2 } \label{z_enc_wolf}
\end{equation}  
which is a slightly rearranged Eq.\ (8) from Ref.\
\onlinecite{encounter_model0} written in our notation. In the
thermodynamic limit $N\to\infty$ $W^{enc}$ can be found from the continuum
inverse Fourier transform as
\begin{equation}
	W^{enc}_{\bf l}=\frac{1}{(2\pi)^{3}}
	\iiint_{-\pi}^\pi
	d\bar{{\bf K}}\,W^{enc}(\bar{{\bf K}})e^{-i{\bf l}\cdot\bar{{\bf K}}}
	\label{W_enc}
\end{equation}
where dimensionless momentum $\bar{\bf K}=a{\bf K}/2$ was introduced
to simplify notation and the use has been made of the cubic
symmetry. Substituting this into Eq.\ (\ref{z_enc_wolf}) and using
the identity
\begin{equation}
	\delta(\bar{{\bf K}})=\frac{1}{(2\pi)^3}\sum_{{\bf l}}
	e^{-i{\bf l}\cdot\bar{{\bf K}}}
	\label{delta-function}
\end{equation}
it can be seen that $z_{enc}$ depends only on the behavior of
$W^{enc}({\bf K})$ at small $|{\bf K}|$. Indeed, the sum over
${\bf l}$ in Eq.\ (\ref{z_enc_wolf}) is calculated by first applying
the Laplacian $-\nabla^2_{\bar{\bf K}}$ to Eq.\ (\ref{delta-function})
and then carrying out the integration over ${\bf K}$ by parts twice. The
second derivatives at $\bar{\bf K}=0$ are calculated with the use of Eq.\
(\ref{Wenc-approx}) to give
\begin{equation}
	z_{enc}\simeq\frac{12\lambda^2}{f_\infty a^2}.
	\label{z_enc_lambda}
\end{equation}
where we replaced general $f$ by $f_\infty$ from Eq.\ (\ref{f_infty})
because the phenomenological approach was developed for the case of strong
I-v binding.  Now substituting $\lambda$ from Eq.\ (\ref{lambda}) into Eq.\
(\ref{z_enc_lambda}) and using the definition of $D_m$ Eq.\ (\ref{Dm})
one finds
\begin{equation}
	z_{enc}=\frac{w_2}{r}=\frac{w_2}{7w_3p_\infty}
	\label{z(w2-w4)}
\end{equation}
where use has been made of the definition of $r$ Eq.\ (\ref{r}).
As will be discussed below, this expression does not agree with the
qualitative conclusions about the dependence of $z_{enc}$ on the 5FM
frequencies reached in Ref.\ \onlinecite{voglAlFe} on the basis of
the MC simulations. The comparison may be not warranted because Eq.\
(\ref{z(w2-w4)}) has been derived for the strong-coupling case while
simulations covered also other cases.

To clarify this issue let us calculate $z_{enc}$ for the general 5FM
within the approach of Ref.\ \onlinecite{bender}. In this approach the
mean number of the impurity jumps is calculated as
\begin{equation}
	z_{enc}=1/(1-p_R)
	\label{z_enc}
\end{equation}
where $p_R$ is the probability for the vacancy to return from the first
CS on the impurity site.  $p_R$ can be found as follows. The rate of the
vacancy return on the impurity site is equal to the rate of I-v exchanges
$w_2$ irrespective of the binding. The rate of the definite departure
of the vacancy away from the impurity is equal to the product of the
total rate of the vacancy jumps from the first CS to higher shells $7w_3$
multiplied by the probability to diffuse infinity far from the impurity
$p_\infty$. Each factor in the product $r=7w_3p_\infty(w_4/w_0)$ can
also be calculated for any values of the frequencies, not only for the
strong binding case. With all rates being known, the return probability
is found as the ratio of the return rate to the total rate\cite{n-fold}
\begin{equation}
	p_R=\frac{w_2}{w_2+r}.
	\label{p_R}
\end{equation}
Substituting this into Eq.\ (\ref{z_enc}) one finds
\begin{equation}
	z_{enc}=1+\frac{w_2}{r}=1+\frac{w_2}{7w_3p_\infty}.
	\label{z_enc2}
\end{equation}
This expression differs from Eq.\ (\ref{z(w2-w4)}) only on a numerical
constant equal to unity while the dependence on the frequencies is the
same.  It is interesting to note that being derived in a more general
case Eq.\ (\ref{z_enc2}) looks as more reliable than Eq.\ (\ref{z(w2-w4)})
restricted to the case of tight binding. However, the latter seems to be
more physical because when $w_2$ (hence, $p_R$) vanish it predicts the
correct number of impurity jumps equal to zero while Eq.\ (\ref{z_enc2})
predicts one jump. Nevertheless, below we will use $z_{enc}$ from Eq.\
(\ref{z_enc2}) for consistency with Ref.\ \onlinecite{voglAlFe}.

\begin{figure}
\begin{center}
\includegraphics[viewport = 0 0 200 200, scale = 1.]{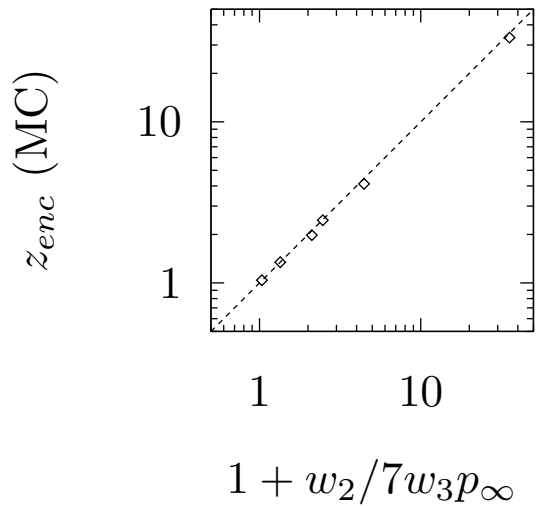}
\end{center}
\caption{\label{fig3-2} The number of impurity jumps $z_{enc}$
during single I-v encounter as obtained in MC simulations in Ref.\
\onlinecite{voglAlFe} and as calculated with Eq.\ (\ref{z_enc2}); the
dashed line corresponds to strict equality between the two values.}
\end{figure}

From Fig.\ \ref{fig3-2} it is seen that $z_{enc}$ from Eq.\ (\ref{z_enc2})
agrees with the MC simulations for all sets of the 5FM parameters studied
in Ref.\ \onlinecite{voglAlFe}. But the qualitative description of the
frequency dependence of $z_{enc}$ suggested in the paper does not agree
with Eq.\ (\ref{z_enc2}).  First disagreement concerns the conclusion
that $z_{enc}$ strongly depends on $w_1/w_3$ ratio and weakly depends
on the ratio $w_4/w_0$.\cite{voglAlFe}  But Eq.\ (\ref{z(w2-w4)}) does
not depend on $w_1$ at all while through $p_\infty$ it strongly depends
on $w_4/w_0$ ratio, especially when it is large, as was shown with the
use of MC simulations in I and will be further confirmed below within
the rigorous solution. Second, when $w_4/w_0\to\infty$ $p_\infty\propto
w_0/w_4$ and so $z_{enc}$ in Eq.\ (\ref{z(w2-w4)}) will also strongly
depend on the $w_4/w_3$ ratio while in Ref.\ \onlinecite{voglAlFe}
it was concluded that $z_{enc}$ does not depend on it.  One source of
the discrepancies seems to be due to the fact that the case of large
$w_4/w_0$ ratios was not investigated in Ref.\ \onlinecite{voglAlFe}
while in our study it was of major importance because the first-principles
calculations discussed in I predict that it dominates strong I-v binding,
at least in the aluminum host.

Another apparent disagreement concerns roughly Gaussian diffusion
profiles obtained in the MC simulations in Ref.\ \onlinecite{voglAlFe}
despite the large I-v attraction which according to our findings might
lead to NGDPs. This, however, is explainable by the small value of
the parameter $\lambda=0.4a$ that can be calculated with the use of
Eqs.\ (\ref{z_enc_lambda}) and (\ref{f_infty}) on the basis of the
experimentally fitted frequencies and $z_{enc}\approx2$.\cite{voglAlFe}
The small values of both quantities are mainly due to the small value
of $w_2$ found in the fit. Thus even in the case of strong binding the
mobility of impurity can be low because of the reduced frequency of
the I-v exchanges that define the diffusion.  $\lambda$ characterizes
the extent of the exponential part of the NGDP. But with $\lambda=0.4a$
being smaller than even the NN distance non-Gaussian behavior is hardly
detectable in the 3D case, though in 2D the microscopic profiles were
reliably measured in Refs.\ \onlinecite{Co/cu(001),Pd/Cu}.  The small
value of $z_{enc}$ and $\lambda$ in the Fe\underline{Al} system were
partly due to the high experimental temperature of 923~K.  There is not
enough data to decide whether NGDPs can be observed in this system at
lower temperature, though $\lambda$ usually grows when the temperature
is lowered.\cite{cowern1990,cowern2003,I}
\section{\label{001}Diffusion along a symmetry axis}
A major goal of the present study is to give a rigorous justification
to the phenomenological theory of impurity diffusion in the FCC
host developed in I. The theory has been based on the notion of
the mobile intermediate state of the impurity introduced in Ref.\
\onlinecite{cowern1990} which in the case of the 5FM has been identified
in I with the tightly bound I-v pairs. This has made possible derivation
of simple formulas for the impurity GF and the diffusion profiles
in stark contrast with the complicated expression for the rigorous
solution of Sec.\ \ref{solution}. So our aim in this section is to
show that the phenomenological expressions do agree with the rigorous
approach in the limits of strong I-v binding and the macroscopic
diffusion that corresponds to small values of $z$ and $|{\bf K}|$ in the
frequency-momentum space or, equivalently, to macroscopic spatiotemporal
scale.

In cubic crystals diffusion is isotropic in the limit of small $|{\bf K}|$
so instead of a general momentum ${\bf K}$ that requires solution of 54
equation, the momentum along a symmetry axis can be chosen in order to
reduce the system size.  To this end let us chose the direction along
$Z$ axis with ${\bf K}=(0,0,K)$. In this case the components of vector
$\vec{\rho}_{\bf K}$ in Eq.\ (\ref{canonical}) can be divided into 13
equivalence classes, as shown in Fig.\ \ref{fig1-2}. All components within
a class are equal so it is sufficient to retain only one equation for each
class which reduces the system of 54 equations to a system of 13 equations
\begin{equation}
	(I-HU)\vec{\rho}_{K}=P\vec{c}_0,
	\label{13x13}
\end{equation}
where on the r.h.s.\ only 13 components of vector Eq.\ (\ref{Prho})
corresponding to different classes has been kept, $I$ is the identity
matrix of size 13 and $H$ and $U$ are $13\times14$ and $14\times13$
matrices, respectively, that are obtained from $P$ and $V$ as follows.
Because all vector components  belonging to one class are the same,
their precise coordinates are irrelevant and can be characterized only
by the class number which is used as the subscripts of matrices $H$
and $U$.  To find $H_{ij}$ one can take any site ${\bf n}$ from class $i$
and then sum $P_{\bf n-n^\prime}$ over all ${\bf n}^\prime$ from class
$j$. For example,
\begin{equation}
	H_{51} = P_{011}  + P_{013}+ 2P_{112},
	\label{H51}
\end{equation}
as can be seen from Fig.\ \ref{fig1-2} by taking the leftmost site
from class 5 and checking that there is exactly one NN site at ${\bf
n}=(1,1,0)$ belonging to class 1 (the leftmost in the figure from this
class), one at ${\bf n}=(3,0,1)$ (the rightmost in the class) and two
sites at ${\bf n}=(2,\pm1,1)$. It is to be noted that matrices $H$ and
$U$ are not symmetric. Besides, one would need also the matrix elements
$H_{i0}$, where 0 refers to the ``class'' consisting of the site at
the impurity position. $H_{i0}$ are necessary to fill the 0-th column
of $13\times14$ matrix $H$. All expressions of the matrix elements of
$H$ in terms of $P_{ijk}$ needed in calculations below can be found in
ancillary file in Ref.\ \onlinecite{SM}.

Matrix $U$ can be obtained from $V$ in a similar way by summing all
contributions $V_{{\bf n,n}^\prime}$ into matrix elements between
the classes:  
\begin{widetext}
\begin{eqnarray}
\label{U}
&&\begin{array}{ccccccccccccc}\hspace{3.7em}1\hspace{2.2em}&2\hspace{1.7em}&3
\hspace{2.em}&4\hspace{2.em}&5\hspace{1.8em}&6\hspace{2.em}&7\hspace{1.9em}&8
\hspace{1.9em}&9\hspace{1.5em}&10\hspace{1.5em}&11\hspace{.9em}&12
\hspace{1.1em}&13
\end{array}\nonumber\\[-1ex]
U=&&
\renewcommand*{\arraystretch}{1.1}
\begin{array}{r}0\\[.05ex]1\\[.05ex]2\\[.05ex]3\\[.05ex]4\\[.05ex]5
\\[.05ex]6\\[.05ex]7\\[.05ex]8\\[.05ex]9\\[.05ex]10\\[.05ex]11\\[.05ex]12
\\[.05ex]13\end{array}
\renewcommand*{\arraystretch}{0.85}
\left[\begin{array}{c|c|c|c|c|c|c|c|c|c|c|c|c}
4\bar{w}_5&4\bar{w}_5&4\bar{w}_5&&&&&&&&&&\\\hline
U_{11}&2\bar{w}_1&{w}_{2}e^{i\bar{K}}&\bar{w}_4&\bar{w}_4&&2\bar{w}_4&2\bar{w}_4&&&\bar{w}_4&&\\\hline 
2\bar{w}_1&U_{22}&2\bar{w}_1&&2\bar{w}_4&&&2\bar{w}_4&2\bar{w}_4&&&\bar{w}_4&\\\hline
{w}_{2}e^{-i\bar{K}}&2\bar{w}_1&U_{11}&&\bar{w}_4&\bar{w}_4&&&2\bar{w}_4&2\bar{w}_4&&&\bar{w}_4\\\hline
4\bar{w}_3&&&-4\bar{w}_4&&&&&&&&&\\\hline
\bar{w}_3&2\bar{w}_3&\bar{w}_3&&-4\bar{w}_4&&&&&&&\\\hline
&&4\bar{w}_3&&&-4\bar{w}_4&&&&&&&\\\hline
2\bar{w}_3&&&&&&-2\bar{w}_4&&&&&&\\\hline
\bar{w}_3&\bar{w}_3&&&&&&-2\bar{w}_4&&&&&\\\hline
&\bar{w}_3&\bar{w}_3&&&&&&-2\bar{w}_4&&&&\\\hline
&&2\bar{w}_3&&&&&&&-2\bar{w}_4&&&\\\hline
\bar{w}_3&&&&&&&&&&-\bar{w}_4&&\\\hline
&\bar{w}_3&&&&&&&&&&-\bar{w}_4&\\\hline
&&\bar{w}_3&&&&&&&&&&-\bar{w}_4
\end{array}   \right], 
\end{eqnarray}
\end{widetext}
where
\begin{eqnarray}
	U_{11}&=&-2\bar{w}_1-w_2-7\bar{w}_3-\bar{w}_5\\
	U_{22}&=&-4\bar{w}_1-7\bar{w}_3-\bar{w}_5.
\end{eqnarray}

Matrix $U$ has the familiar gain-loss structure (see Appendix
\ref{master_eq}) with the barred frequencies $\bar{w}_i$ in the
off-diagonal matrix elements having plus sign and the diagonal terms with
the minus sign, but not necessary positive or negative values because
of the subtraction of $w_0$.  Thus, for example, $U_{41}$ is equal to
$4\bar{w}_3$ because there is exactly four sites that are NN to a site
in 4th class from which a vacancy may jump at this site with the rates
different from $w_0$ and similarly $U_{44}$ is equal to $-4\bar{w}_4$
because there is four sites (all from class 1) where the vacancy can
jump from a site in class 4 (see Fig.\ \ref{fig1-2}).

Because our goal is to find the impurity GF, we are interested only
in those components of vector $\vec{\rho}_{{\bf K}=(0,0,K)}\equiv
\vec{\rho}_K $ that enter into Eq.\ (\ref{zG}), that is, only in the
components belonging to the first and the third classes.  Moreover,
because for any NN vector ${\bf e}^{(1)}$ from the first class there
exists an NN vector in the third class such that
\begin{equation}
{\bf e}^{(3)}=-{\bf e}^{(1)},
\label{e3=-e1}
\end{equation}
it is easily seen that one can express the solution for $G(K,z)$
in terms of only one of the components because the FCC lattice is
centrosymmetric.  Because the initial value Eq.\ (\ref{ini}) was also
assumed to be centrosymmetric, the kinetics will preserve the symmetry,
so during the whole system evolution the following equality will hold
\begin{equation}
	\langle \hat{\imath}_{\bf-l}\hat{v}_{\bf-l-n}\rangle=
	\langle \hat{\imath}_{\bf l}\hat{v}_{\bf l+n}\rangle. 
\label{r2-r}
\end{equation}
Applying the LF-transform Eq.\ (\ref{LFdef}) to this equation and 
changing the summation on the r.h.s.\ from ${\bf K}$ to $-{\bf K}$ 
one arrives at the equality
\begin{equation}
\rho({\bf K},z,-{\bf n})=\rho({\bf-K},z,{\bf n}).
\label{n2-n}
\end{equation}
Thus, in view of Eq.\ (\ref{e3=-e1}), only two functions $\rho_{\pm
K}(z,{\bf e}^{(k)})$ with $k$ being either 1 or 3 are needed for the
calculation of the impurity GF in Eq.\ (\ref{zG}) .  Choosing $k=1$
with the use of the Cramer's rule one gets
\begin{equation}
\rho(K,z,{\bf e}^{(1)})=\Delta_1(K,z)/\Delta(K,z),
\label{cramers}
\end{equation}
where 
\begin{equation}
	\Delta(K,z) =\det(I-HU)
	\label{det0}
\end{equation}
is the determinant of the system in Eq.\ (\ref{13x13}) and $\Delta_1(K,z)$
is this determinant with the first column replaced by the r.h.s.\ vector
$\tilde{P}\vec{c}_0$ Eq.\ (\ref{dPc}) according to the Cramer's rule.
Due to the axial symmetry Eq.\ (\ref{SzK}) simplifies to
\begin{equation}
	\Sigma(K,z)=-w_2a^2
	\sum_{\left\{\parbox{1.5em}{\tiny upper lower}\right\}}(1-e^{\mp ia{K}/2})
z\rho({\pm K},z,{\bf e}^{(1)}),
	\label{DzK_Z}
\end{equation}
where the summation is over the upper and the lower signs in the
summand and ${\bf e}^{(1)}$ is any of the four lattice vectors from
the first class.  Eq.\ (\ref{DzK_Z}) together with Eq.\ (\ref{dyson})
solves the problem of finding the impurity GF in the uniaxial geometry.
\section{\label{D-r-Dm}Diffusion of tightly bound I-v pairs}
In the axisymmetric case the impurity GF is fully determined by the diffusion
kernel $\Sigma$ Eq.\ (\ref{DzK_Z}) which in the limit $K\to0$ in FCC lattice 
can also describe macroscopic diffusion for any ${\bf K}$ with small absolute
value. 

In the phenomenological theory only the macroscopic diffusion can be
treated. The corresponding diffusion kernel can be found from Eqs.\
(44) and (53) in I as
\begin{equation}
\Sigma^{(ph)}(K,z)=-\frac{12c_{NN}z+g}{z+r+D_mK^2}D_mK^2,
	\label{ph-kernel}
\end{equation}
where ${\bf K}^2$ was replaced by $K^2$ to facilitate comparison with
the above rigorous expression.  Thus, in order to prove equivalence of
the phenomenological approach and the rigorous theory in the diffusion
limit in the case of strong I-v binding it is necessary to show that Eq.\
(\ref{ph-kernel}) approximately reproduces Eq.\ (\ref{DzK_Z}) at small
values of $z$, $K$, and $w_3/w_4$ (see Eq.\ (\ref{w3/w4})).

To this end let us first compare the values of the diffusion constant
in both approaches. Using Eq.\ (\ref{dyson}) it is easy to
show that the diffusion constant is obtained from the diffusion kernel as
\begin{equation}
D=\lim_{z\to0,K\to0}\Sigma(K,z)/K^2\equiv\lim_{z\to0,K\to0}{\cal D}(K,z).
	\label{DDef}
\end{equation}
Applying this to Eq.\ (\ref{ph-kernel}) one gets
\begin{equation}
	D_\infty=gD_m/r,
	\label{DD}
\end{equation}
where the diffusion constant was supplied by the subscript $\infty$ to
remind that the phenomenological expression is valid only in the case of
strong I-v binding. Eq.\ (\ref{DD}) is satisfied in the phenomenological
theory (see Eq.\ (55) in I) and we are going to show that this is also
the case in the rigorous approach. This amounts to showing that $D_\infty$
agrees with the known expressions for $D$ in the tight-binding limit. But
this would also hold if $D$ is correct for all values of frequencies,
not only in the limiting case. Though rigorous analytic expression
for $D$ in the canonical version of the 5FM has been derived in Ref.\
\onlinecite{koiwa1983}, below we present its numerical calculation which
has an advantage of admitting generalization to the extended versions
of the model.\cite{LECLAIRE1978,wu_high-throughput_2016,Bocquet,NNN_jumps}
\subsection{\label{diffusion-const}The diffusion constant}
Correlation effects are conventionally accounted for in the
diffusion constant through the correlation factor $f$ defined
as\cite{LECLAIRE1978,manning,philibert}
\begin{equation}
D=c_vw_2\frac{w_4}{w_3}fa^2,
\label{D}
\end{equation}
where $f$ in the 5FM case was shown to have the
general form\cite{manning,koiwa1983,Bocquet}
\begin{equation}
f=\frac{2w_1+7F(w_4/w_0)w_3}{2w_1+2w_2+7F(w_4/w_0)w_3}.
\label{f}
\end{equation}
This expression has been thoroughly studied in literature
and so may serve as a stringent test for new calculation
techniques.

The limit $K\to0$ in Eq.\ (\ref{DDef}) can be calculated by first
expanding the $K$-dependent quantities in the nominator and denominator
of Eq.\ (\ref{DzK_Z}) to the second order in $K$ as
\begin{equation}
	{\cal D}(0,z)=w_2a^2z\rho(0,z,{\bf e}^{(1)})
	+\left. 4w_2ia\frac{d[z\rho(K,z,{\bf e}^{(1)})]}{dK}\right|_{K=0}.
	\label{D(z0)}
\end{equation}
In this notation
\begin{equation}
D=\lim_{z\to0} {\cal D}(0,z).
\label{DcalD}
\end{equation}
The limit is straightforward to calculate for the first term on the
r.h.s.\ of Eq.\ (\ref{D(z0)}).  As is easily seen from the definition of
the LF transform Eq.\ (\ref{LFdef}), if the correlation function Eq.\
(\ref{I-V}) has a non-zero asymptotic when $t\to\infty$, the Laplace
part of the transform develops the singularity $\sim 1/z$.  But at large
times and in the absence of macroscopic fluxes, which is the case when
$K=0$, all kinetics die out and the correlation function tends to its
equilibrium value
\begin{equation}
\langle \hat{\imath}_{\bf l}\hat{v}_{\bf l+n}\rangle|_{t\to\infty} 
\to N^{-1}\langle \hat{\imath}_{\bf 0}\hat{v}_{\bf n}\rangle ^{(eq)}
=N^{-1}c_{\bf n}^{(eq)},
	\label{equilibrium}
\end{equation}
where $c_{\bf n}^{(eq)}$ is the equilibrium vacancy concentration at
site ${\bf n}$ relative to the impurity; the factor $1/N$ appears due to
the fact that as $t\to\infty$ the only impurity in the system occupies
any of $N$ sites with equal probability $1/N$.  The LF transform Eq.\
(\ref{LFdef}) of Eq.\ (\ref{equilibrium}) at ${\bf K=0}$ gives
\begin{equation}
	\lim_{z\to0}z\rho(0,z,{\bf n})=c_{\bf n}^{(eq)}.
	\label{limz20eq}
\end{equation}
Thus, the first term on the r.h.s.\ of Eq.\ (\ref{D(z0)}) in the limit
$z\to0$ is equal to
\begin{equation}
w_2a^2c_{NN}^{(eq)}=w_2a^2c_vw_4/w_3
\label{1st-term}
\end{equation} 
where use have been made of Eqs.\ (\ref{c^eq}) and (\ref{w3/w4}).  

\begin{figure}
\begin{center}
\includegraphics[viewport = 212 340 450 650, scale = 0.6]{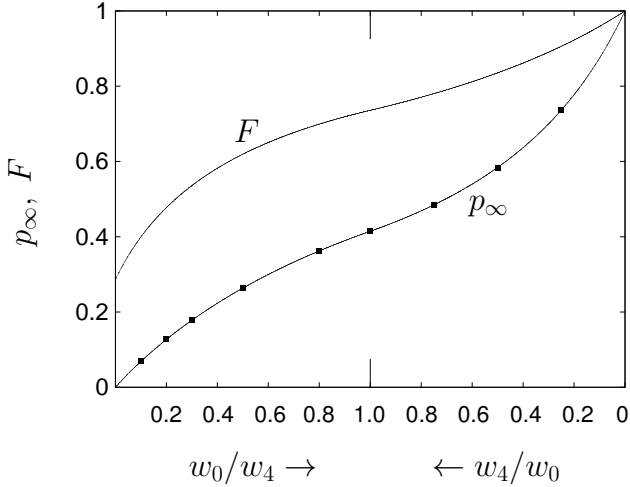}
\end{center}
\caption{\label{fig4-2}Upper curve: $10^4$ points for $F$ from Eq.\
(\ref{F}) calculated for random values of ratios $w_i/w_0$, $i=1-4$;
on the scale of the figure the data would be indistinguishable from the
curves of Refs.\ \onlinecite{manning,koiwa1983}. Lower curve: $10^4$
points $p_\infty$ calculated with the use of Eq.\ (\ref{r5fm}) for random
ratios $w_i/w_0$, $i=1, 2, 4$, $w_3=0$; symbols: MC simulations from I.}
\end{figure}

The limit $z\to0$ in the last term of Eq.\ (\ref{D(z0)}) concerns only
$\rho$ so it is convenient to first obtain it with the use of Eq.\
(\ref{Prho})
\begin{equation}
\lim_{z\to0}zP\vec{c}_0=c_v 
\label{limzPc}
\end{equation}
and (\ref{cramers}) as
\begin{equation}
	\lim_{z\to0}z\rho({K},z,{\bf e}^{(1)})=	
	c_v\bar{\Delta}_1(K,0)/\Delta(K,0),
	\label{lim_z}
\end{equation}
where $\bar{\Delta}_1$ differs from ${\Delta}_1$ in that instead of
$P\vec{c}_0$ the limiting value Eq.\ (\ref{limzPc}) is substituted and,
besides, the vacancy concentration is factored out of the determinant
so that the first column of $\Delta(K,0)$ should be replaced simply by
the column of unities.  Further, using the reflection symmetry of the
1D geometry it can be shown that determinant $\Delta(K,0)$ is an even
function of $K$, so only $\bar{\Delta}^1$ contributes to the derivative
in Eq.\ (\ref{D(z0)}). The dependence of $\bar{\Delta}^1$ on $K$ comes
only from the matrix element
\begin{equation}
	U_{13}=w_2\exp(iaK/2)	
	\label{u_13}
\end{equation}
of matrix Eq.\ (\ref{U}) that contributes to the third column of matrix
$I-HU$ the terms of the form:
\begin{equation}
	\delta (I-HU)_{j3}=-w_2H_{j1}\exp(iaK/2).
	\label{delta_j3}
\end{equation}
Similar contributions to column one from the matrix element $U_{31}$ disappear 
from $\bar{\Delta}_1$ because this column is replaced by unities. Thus,
differentiation of $\bar{\Delta}^1$ with respect to $K$ in the limit $K\to0$
amounts to the following expression for the second term in the diffusion
constant Eq.\ (\ref{D(z0)})
\begin{equation}
\left. 4w_2ia\frac{d[z\rho(K,z,{\bf e}^{(1)})]}{dK}\right|_{z,K=0}
=\left.2c_vw_2a^2 \frac{\bar{\Delta}_{13}}{\Delta}\right|_{z,K=0},
	\label{2nd-term}
\end{equation}
where $\Delta_{13}$ is the determinant of matrix $I-HU$ where the
first column is replaced with unities and the third one with the terms
$w_2H_{j1}$. Now adding the two terms in Eq.\ (\ref{D(z0)}) in the
limit $z\to0$ with the use of Eqs.\ (\ref{limz20eq}), (\ref{w3/w4}), and
(\ref{2nd-term}) and using the definition of the correlation factor Eq.\
(\ref{D}) one arrives at
\begin{equation}
f=1+2\frac{w_3}{w_4}\left.\frac{\bar{\Delta}_{13}}{\Delta}\right|_{z,K=0}.
	\label{f_calc}
\end{equation}
To compare with $f$ in Eq.\ (\ref{f}) the latter was solved with 
respect to $F$ as
\begin{equation}
F\left(\frac{w_4}{w_0}\right)=\frac{2}{7w_3}\left(\frac{fw_2}{1-f}-w_1\right)
	\label{F}
\end{equation}
and $f$ from Eq.\ (\ref{f_calc}) was calculated for more than $10^5$
randomly generated quartets of the ratios $10^{-5}<w_i/w_0<10^{5}$,
$i=1-4$. The calculated values were substituted in Eq.\ (\ref{F}) and as
can be seen from Fig.\ \ref{fig4-2}, where for brevity only about
10\% of the simulation data are presented, no noticeable dependence on
$w_i/w_0$, $i=1-3$ is discernible.  This supports the results of Refs.\
\onlinecite{manning,koiwa1983} and confirms the assumption that the
general form of $f$ in Eq.\ (\ref{f}) is exact to the leading order
in $c_v$.
\subsection{\label{I-v}Diffusion kernel in the strong-binding limit}
The expression for $D$ obtained from Eqs.\ (\ref{D}) and (\ref{f_calc}) is
valid for all values of the frequencies, including the tight-binding case.
Thus, Eq.\ (\ref{DD}) provides one relation between three phenomenological
parameters and the known diffusion constant so in order to calculate all of
them in the rigorous approach two additional relations are needed.  Let us
begin with the calculation of the decay rate $r$.  It does not enter as a
parameter in the rigorous solution but has to be found from the assumption
that in the tight-binding case the exact diffusion kernel has the form of
Eq.\ (\ref{ph-kernel}) and that $r$ depends only on $w_3$ and $w_4/w_0$
as in the phenomenological expression Eq.\ (\ref{r}).  Confirmation of
these assumptions in the framework of the rigorous theory would provide
a nontrivial check of the phenomenological approach.

As is seen from Eq.\ (\ref{w3/w4}), strong I-v binding means either
small value of $w_3$, or large value of $w_4$, or both. The difficulty in
studying the tight-binding limit within the rigorous approach is that the
cases $w_3\to0$ and $w_4\to\infty$ should be studied separately because
the frequencies enter in a different manner in both the decay rate Eq.\
(\ref{r}) and in the determinants that define the diffusion kernel, as
can be seen from Eqs.\ (\ref{U}), (\ref{cramers}), (\ref{13x13}), and
(\ref{SzK}). Therefore, let us start from the $w_3\to0$ case. We first
note that the determinants in Eq.\ (\ref{ph-kernel}) are polynomial
in their matrix elements so the singular behavior at $z,r,K\to0$
may originate only from the determinant in the denominator of Eq.\
(\ref{cramers}) which in this limit should be representable as 
\begin{equation}
\Delta = \Psi(z,K,w_3)(z+\underline{r}+\underline{D_m}K^2),
\label{product}
\end{equation}
where $\underline{r}$ and $\underline{D_m}$ are underlined to
indicate that in this expressions they in general are functions of $z$
and $K$. When the latter variables are equal to zero the underlined
quantities would coincide with the constants $r$ and $D_m$.  The function
$\Psi$ is fixed by the condition that $z$ in the second factor has the
coefficient equal to unity when $z$, $K$, and $w_3$ are equal to zero and
$\Psi(0,0,0)\not=0$; for brevity, only the variables that change their
values in the expressions below are explicitly shown as the arguments
of $\Psi$ with other parameters being treated as constants.

From Eq.\ (\ref{product}) one can derive the derivatives of the determinant
over its arguments as
\begin{equation}
\partial \Delta/\partial z|_{z,w_3,K=0}=\Psi(0,0,0), 
\label{Delta_prime}
\end{equation}
(additionally assuming that $\underline{r}\propto w_3$) so that if the
decay rate has the form of Eq.\ (\ref{r}) then
\begin{equation}
	p_\infty = \frac{1}{7}\left.\frac{\partial \Delta/\partial w_3}
	{\partial \Delta/\partial z}\right|_{z,K,w_3=0}.
	\label{r5fm}
\end{equation}
should be the function of only the ratio $w_4/w_0$ and should agree
with the runaway probability defined in I. In Fig.\ \ref{fig4-2}
it can be seen that this is indeed the case.

Similarly, 
\begin{equation}
	D_m=\left.\frac{\partial \Delta/\partial K^2}
	{\partial \Delta/\partial z}\right|_{z,K,w_3=0}.
	\label{Dm5fm}
\end{equation}
calculated for $\sim10^5$ randomly generated frequency ratios agreed with
Eq.\ (\ref{Dm}) to within the accuracy of the calculations. With the use
of Eq.\ (\ref{DD}) and taking into account that $D$, $D_m$, and $r$ in
the tight-binding case have the same values as in the phenomenological
theory to the accuracy of the numerical calculations, we conclude that
the value of the association rate $g$ in the rigorous approach also has
the correct value.

To finalize comparison with the phenomenological theory in the $w_3\to0$
case it remained to consider the contribution due to the associated pairs
that may be present in the initial state. The contribution is given by the 
term in Eq.\ (\ref{ph-kernel}) proportional to $c_{NN}$
\begin{equation}
\delta\Sigma^{(ph)}=-\frac{12c_{NN}z}{z+r+D_mK^2}D_mK^2.
\label{dSigma}
\end{equation}
In Eq.\ (\ref{DzK_Z}) this term should originate from Eq. (\ref{cramers})
through the similar contribution in the initial condition Eq.\ (\ref{Prho})
\begin{equation}
	\delta(\tilde{P}\vec{c}_{0})_{\bf n}=12c_{NN}P_{\bf n}.
\label{dPc}
\end{equation}
In the Cramer's rule solution it can be separated from the rest of the
terms because determinants are linear functions of the column vectors.
Explicit expression for $\delta\Sigma$ can be derived from Eqs.\
(\ref{DzK_Z}) and (\ref{cramers}) with $\Delta_1$ in the last equation
replaced by $\Delta_{\vec{P}}$, where $\vec{P}$ is the vector of size 13
composed of $P_{\bf n}$ from Eq.\ (\ref{dPc}) with ${\bf n}$ belonging
to the respective classes (all $P_{\bf n}$ within the class are the same).
Now taking into account representation Eq.\ (\ref{product}) for $\Delta$,
$\delta\Sigma$ from Eq.\ (\ref{DzK_Z}) can be cast in the form similar
to Eq.\ (\ref{dSigma}) as
\begin{equation}
\delta\Sigma\simeq-\theta(z,K,w_3)\frac{12c_{NN}z}{z+\underline{r}
+\underline{D_m}K^2}\underline{D_m}K^2
\label{dSigma-xct}
\end{equation}
where notation is the same as in Eq.\ (\ref{product}) and in
$\theta(z,K,w_3)$ are gathered all factors not present in the rest of
the equality.  Next, from Eq.\ (\ref{dSigma}) it is easy to see that
\begin{equation}
-\lim_{K\to0}\left[{\delta\Sigma^{(ph)}}/(12c_{NN}z)\right]_{z,r=0}=1.
\label{dSigma=1}
\end{equation}
Similar transformation of Eq.\ (\ref{dSigma-xct}) leads to the
following condition of the agreement between the rigorous approach and
the phenomenological theory:
\begin{equation}
\theta(z\to0,K\to0,w_3\to0)\to1.
\label{theta=1}
\end{equation}
The limit $K\to0$ in the expression for $\delta\Sigma$ in terms of the
determinant ratio can be found similar to the case of the diffusion
constant in Sec.\ \ref{diffusion-const}. But a simpler route is to
calculate it numerically.  For $\bar{K}=0.01$ $10^5$ trios of the
ratios $w_1/w_0$, $w_2/w_0$, and $w_4/w_0$ has been generated and Eq.\
(\ref{theta=1}) has been found to be satisfied with very high accuracy.

In the second case of tight-binding $w_4/w_0\to\infty$ all
calculations were performed along the above lines with a few technical
modifications. From the computational point of view the main difference
between the two cases is that the determinants in Eq.\ (\ref{cramers}) are
polynomials of tenth order in $w_4$, as can be seen from Eqs. (\ref{U})
and (\ref{det0}) so when $w_4\to\infty$ their ratio may be hard to compute
with sufficient precision. The difficulty was resolved by dividing the
last ten columns in the determinants by $w_4$ so both the matrix elements
and the determinants became bounded as $w_4\to\infty$ and thus easily
manageable numerically. The condition $w_3=0$ in the equations above
was replaced by the condition $1/w_4=0$ and the derivative over $w_3$
replaced by the derivative with respect to $1/w_4$. The calculations
performed with the use of this trick confirmed asymptotic equivalence
of the rigorous solution to the phenomenological theory also in this case.
\section{\label{general_case}I-v interaction of arbitrary strength}
The phenomenological theory developed in I has been based on the notion
of the mobile impurity state which in the case of the 5FM has been
identified with the tightly bound I-v pairs.\cite{cowern1990,In/cu(001),I}
In previous sections we succeeded in showing that the expressions derived
in I can be rigorously justified within the rigorous approach but only in
the limit of tight binding. But this regime is obviously not universal and
in different systems I-v interaction will exhibit either only a weak or no
attraction\cite{Fe57Cu} or the repulsion.\cite{wu_high-throughput_2016}
The rigorous solution in principle covers all possible cases
but in view of the physical transparency and simplicity of the
phenomenological theory it would be desirable to be able to asses its
applicability in the cases when the tight-binding condition is not
obviously satisfied.\cite{voglAlFe} Besides, the pair diffusion picture
predicts such interesting and unusual phenomena as the NGDPs and the
non-Fickian diffusion but as was noted in the Introduction, in Refs.\
\onlinecite{Brummelhuis1988,toroczkai1997,Co/cu(001)} it was shown that
NGDPs can occur even in the absence of I-v attraction.  This poses the
question of whether the phenomenological theory does not miss possible
additional contributions into the NGDP phenomenon.  This question should
also be addressed within the rigorous approach.

\begin{figure}
\begin{center}
\includegraphics[viewport = 0 0 400 120, scale = 1.6]{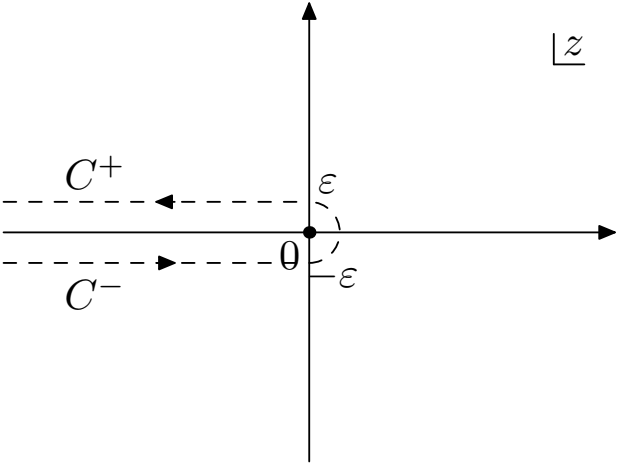}
\end{center}
\caption{\label{fig5-2}Complex-$z$ plane with the deformed contour of
integration over $z$ in the inverse LF transform Eq.\ (\ref{LF-1def})
shown by dashed line. For farther explanations see the text.}
\end{figure}

Unfortunately, the rigorous solution is difficult to deal with because of
its complexity. Even in the simpler uniaxial case it has been expressed
through complex $13\times13$ determinants which apparently can be
calculated only numerically. Besides, to find the diffusion profiles
the solution has to be inversely LF-transformed back to the space-time
variables which is a nontrivial task even in the simple 1D geometry.
Explicit expression for the transformed impurity GF reads:
\begin{equation}
	G_l(t)=\frac{1}{2\pi i}\oint dz\,e^{zt}
	\frac{1}{L}\sum_{K}e^{ia{Kl}/2}G(K,z),
	\label{LF-1def}
\end{equation}
where $L$ is the number of planes along the chosen $\langle100\rangle$
direction. The contour in the integration over $z$ in Eq.\ (\ref{LF-1def})
is defined as the vertical line passing to the right of all singularities
of the integrand.\cite{bromwich} But it can be shown that in the 5FM
all perturbations relax to the equilibrium state corresponding to the
zero-eigenvalue eigenmode and all other eigenvalues are negative real
numbers.\cite{van_kampen,me,cond-mat/0505019} Therefore, the integration
contour can be deformed as shown in Fig.\ \ref{fig5-2}.  Now using the
Green's functions property
\begin{equation}
G(K,z^*)=G^*(K,z)
\label{herglotz}
\end{equation}
(the star means complex conjugate) we can restrict the integration
only to one of the contour branches as
\begin{eqnarray}
\frac{1}{2\pi i}\oint dz\,&&e^{zt}G(K,z) = \frac{1}{\pi}\mbox{Im} 
\int_{C^-} dz\,e^{zt}G(K,z)\nonumber\\
\xrightarrow[\varepsilon\to0]{}&&
\frac{1}{\pi}\int_{-\infty}^0 dE\,e^{Et}\mbox{Im}\,G(K,E^-),
\label{intE}
\end{eqnarray}
where in the limit $\varepsilon\to0$ the integration is over the
negative real axis with $E=\mbox{Re}\,z$ and $E^-=E-i\varepsilon$.

Thus, to calculate the diffusion profile at time $t$ one has to first
integrate the imaginary part over $E$ for every Fourier momentum
$K$ and subsequently recover $G(X=al/2,t)$ via the inverse Fourier
transform. Even for small diffusion length of $\sim10\mbox{ nm}$ in
aluminum where $a/2\approx0.2$~nm one will have to cover the spatial range
$\sim100\;\mbox{nm}\;\simeq500$ lattice planes in $\langle100\rangle$
direction which gives an order of magnitude estimate of the number of
$K$-points (hence, integrations over $E$) needed in the calculation.
Farther, in performing the integrations over $E$ one has to calculate at
every mesh point the complex-valued functions $P_{\bf n}(z)$ from Eq.\
(\ref{P}) that enter into the determinants via matrix $H$ and vector
$P\vec{c}_0$ which can also be calculated only numerically.  And the
mesh in $E$ should be sufficiently fine because, as will we seen below,
$\mbox{Im}G(K,E^-)$ may vary rather steeply. The problem somewhat
alleviates in the diffusion limit $t\gg1/w_0$ when the exponential
function in Eq.\ (\ref{intE}) dumps the integrand at large negative values
of $E$ so the integration range in the vicinity of $E=0$ can be chosen
to be of $O(1/t)\simeq w_0$.  The number of $P_{ijk}(z)$ that need be
calculated in the general case is sixteen (see Ref.\ \onlinecite{SM})
though they can be recursively calculated in terms of only three of them
which, in their turn, can be expressed through the complete elliptic
integrals.\cite{GreenFCC,morita1975,Joyce2011}
\subsection{\label{diss}Dissociation of NN I-v pairs}
Though the bound pairs dominate the diffusion only at sufficiently strong
I-v attraction, the diffusion in the 5FM always goes via exchanges within
NN I-v pairs.  So the difference with the binding case is only in the
availability of the vacancy strongly enhanced in the case of tight
binding. This may amount to larger number of I-v exchanges $z_{enc}$
and to longer mean diffusion distance $\lambda$ during one encounter
which could translate in enhanced diffusion constant and macroscopic
NGDP. However, the example of the Fe\underline{Al} system shows
that this is not always the case. Because of the small value of the
exchange frequency $w_2$ both $z_{enc}\simeq2$ and $\lambda\simeq0.4a$
remain small despite the small decay rate $r$ in the denominator of Eq.\
(\ref{z_enc2}).\cite{voglAlFe} Thus, a question arises on whether there
exists a possibility to detect strong I-v attraction in similar cases.
Below with the use of the rigorous solution we are going to show that
though the transition between the cases of strong binding and its absence
is not abrupt, there exists a qualitative differences between the time
evolution of the bound and non-bound I-v pairs with intermediate cases
exhibiting mixed behavior.

To see this, let us first consider the phenomenological theory where the
diffusion of the bound pair can be described by the LF-transformed pair GF
\begin{equation}
	G_p^{(ph)}({\bf K},z)=\frac{1}{z+r+D_m{\bf K}^2}
	\label{GpKz}
\end{equation}
that in the continuum space satisfies the initial condition 
\begin{equation}
	G_p^{(ph)}({\bf R},t=0)=\delta({\bf R})
	\label{G(t=0)}
\end{equation} 
(see Eqs.\ (35) and (37) in I). As follows from the Fourier transform
definition and from Eq.\ (\ref{intE}), the time evolution of the total
density of bound pairs can be found from the inverse Laplace transform
of the GF at ${\bf K=0}$ as
\begin{equation} 
G_p^{tot}(t)=\frac{1}{\pi}\int_{-\infty}^0dE\,\mbox{Im}\,
G_p({\bf 0},E^-)e^{Et}\overset{(ph)}{=}e^{-rt}, 
\label{e-rt} 
\end{equation} 
where the last equality was obtained with the use of the identity
\begin{equation}
\mbox{Im}\,G_p^{(ph)}({\bf K=0},z=E-i\varepsilon)|_{\varepsilon\to0}
=\pi\delta(E+r).
\label{ImG}
\end{equation}

In the lattice case the definition of the pair GF is not completely
straightforward because NN I-v pair cannot be placed on one site but
occupies two sites.  The point pair of the phenomenological approach is
natural to associate with the smallest pair occupying two NN sites. But
because there are 12 such sites around the impurity on the FCC lattice,
the total density of NN pairs should be equal to the sum of 12 densities
from Eqs.\ (\ref{I-V}) and/or (\ref{directLF}) with ${\bf n=e}$.  In the
${\bf K=0}$ case in Eq.\ (\ref{e-rt}) all 12 densities are equal so the
total pair density
\begin{equation}
G_p({\bf K=0},z)=12\rho({\bf K=0},z,{\bf e})|_{c_v=0,12c_{NN}=1}
\label{G=rho}
\end{equation}
can be calculated with any of the NN vectors ${\bf e}$. In this definition
the vacancy concentrations in the initial state Eq.\ (\ref{ini})
were chosen to correspond to only one NN vacancy near the impurity
in otherwise empty system. Physically this describes evolution of one
associated pair from the initial configuration during time interval when
new associations with the bulk vacancies can be neglected ($c_v=0$). This
may correspond to an early stage of out-of-equilibrium evolution when
$t\ll t_a$ and the number of NN pairs in the initial state is large.
The latter can be achieved by means of an appropriate preparation method,
such as the ion implantation technique,\cite{rmp} or in the tight-binding
as a consequence of the strong I-v attraction when the initial state
is prepared in equilibrium at different temperature and free vacancies
anneal faster than the bound ones.

The density $\rho$ on the r.h.s.\ of Eq.\ (\ref{G=rho}) can be found
with the use of the solution Eq.\ (\ref{cramers}) as
\begin{equation}
	G_p({\bf K=0},z)=12\Delta_1(0,z)/\Delta(0,z)|_{c_v=0,12c_{NN}=1}.
	\label{G=rho2}
\end{equation}
The time evolution of the NN density can be found by substituting
$G_p({\bf K=0},E^-)$ into Eq.\ (\ref{e-rt}) and calculating the integral
over $E$. Unfortunately, in general case the GF can be calculated only
numerically. But a qualitative insight can be obtained from the analytic
expression for the 4FM GF (see Appendix \ref{4FM}) which will be used
below in the discussion of the 5FM.

From Eq.\ (\ref{Prho}) with the vacancy concentrations as in Eq.\
(\ref{G=rho2}) and ${\bf n=e}$ with the use of Eqs.\ (\ref{det_S1-2})
and (\ref{DET4fm}) one arrives at the expression
\begin{equation}
G_p^{4FM}({\bf K=0},z)=\frac{1}{z+\underline{r}}
\label{Gp4fm}
\end{equation}
that formally coincides with its phenomenological homologue Eq.\
(\ref{GpKz}) except that now the decay rate is not a constant but a
function
\begin{equation}
\underline{r}(w_3,z)=w_3\frac{P_{\bf e}^{-1}(z)-z(z+13)}{z+12},
\label{r_}
\end{equation}
where $H_{10}$ was replaced by $P_{\bf e}$.\cite{SM} 

A comment is in order about the dimensionalities in Eq.\ (\ref{r_})
and in equations below. The expressions needed in derivation of the
analytic solution in Appendix \ref{4FM} are rather cumbersome so it
was found convenient to set $w_0=1$ to simplify them. This can be
achieved either by introducing the ``natural'' time unit $w_0^{-1}$
or by restoring the missing dimensional coefficients as powers of $w_0$
in the final expressions. This should not cause difficulty because all
relevant quantities have dimensionality of time or its inverse (the
rate or frequency) while the quantities of dimension of length enter
only via the dimensionless combination $\bar{K}=aK/2$.

\begin{figure}
\begin{center}
\includegraphics[viewport = 0 20 400 393, scale = 0.6]{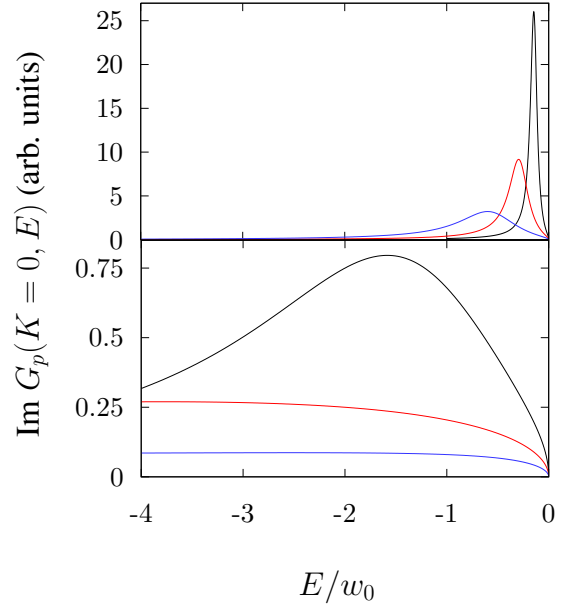}
\end{center}
\caption{\label{fig6-2}(Color online) Imaginary part of $G_p$ as
defined in Eqs.\ (\ref{Gp4fm}) and (\ref{r_}) at the negative real
values of $z$.  In both panels the values of $w_3$ in units of $w_0$ are
0.05, 0.1, 0.2, 0.5, 1 and 2 in order from the largest to the smallest
curves maximums (note difference in scale). For further explanations
see the text.}
\end{figure}

Eq.\ (\ref{relations}) connects $P_{\bf e}(z)$ in Eq.\ (\ref{r_}) with
the extended Watson's integral $P_{\bf 0}(z)$ which analytic properties
are well known (see Ref.\ \onlinecite{Joyce2011} and references therein).
The most important to us is that $P_{\bf 0}(z)$, hence, $P_{\bf e}(z)$
have the square root singularities at $z=0$ (in 3D lattices
to which the discussion below will be restricted) which engenders
important consequences.  Namely, because for $E<0$ $\sqrt{E}$ is imaginary,
the imaginary part of GF in Eq.\ (\ref{Gp4fm}) extends over all
negative values of $E$, as can be seen in Fig.\ \ref{fig6-2} and is
not concentrated at a single point as in the phenomenological approach.
Due to the exponential term in the integrand the $t\to\infty$ asymptotic
of the integral in Eq.\ (\ref{e-rt}) is dominated by the small values of
$|E|$, so the asymptotic will be of the power-law type $\sim t^{-1/2}$,
not exponential as in the phenomenological theory. Moreover, because
for small absolute values of $z$ $|z|\ll\sqrt{|z|}$, {\em a priory} it
is the square-root contribution that should be kept in approximations
for GF at small $|z|$ while $O(z)$ terms should be neglected.  But $z$
in the denominator of $G_p^{(ph)}$ originates from the time derivative of
the phenomenological diffusion equation (see Eq.\ (34) in I and similar
equations in Refs.\ \onlinecite{cowern1990,In/cu(001)}), so by neglecting
it the possibility to describe the I-v pairs as diffusing quasiparticles
is lost.

These observations are fully confirmed by the curves in the lower
panel of Fig.\ \ref{fig6-2} which cannot be approximated by the
sharp delta-function imaginary part of the phenomenological GF in Eq.\
(\ref{ImG}) even crudely. In the upper panel, however, the tallest
curve is strongly peaked and its approximation by a delta-function seems
reasonable.  The difference between the two cases is in the structure
of the denominator in Eq.\ (\ref{Gp4fm}). At small $|E|$ the GF can be
approximated to three leading powers of $E$ as
\begin{equation}
	G_p({\bf K=0},E<0)\approx\frac{1}{E+r(1-iC_r\sqrt{|E|})},
	\label{GpE}
\end{equation}
where $C_r$ is a positive real constant of order unity. We dropped
the 4FM superscript because in Sec.\ \ref{I-v} the validity of Eq.\
(\ref{GpE}) was, in fact, proved in the general 5FM case, though only by
numerical means. The key feature of Eq.\ (\ref{GpE}) is that as $r\to0$,
the square root term enters into GF only with the prefactor $r$, similar
to the exact 4FM expression Eq.\ (\ref{r_}). If it were not so in the
5FM case, the partial derivative over $z$ in Eq.\ (\ref{Delta_prime})
would have been infinite at $z,r=0$ so $p_\infty$ and $D_m$ in Eqs.\
(\ref{r5fm}) and Eq.\ (\ref{Dm5fm}) would be equal to zero, which was not
seen neither in $w_3=0$ case, nor for $w_4=\infty$.  The importance of
the prefactor $r$ is that it makes possible for the phenomenological
theory to hold for the values of $E$ satisfying the inequalities
\begin{equation}
	C_rr\sqrt{|E|}\ll|E|\ll1
	\label{ll_ll}
\end{equation}
which can be satisfied for sufficiently small $r$. Thus, according to Eq.\
(\ref{GpE}) the pole singularity which in the phenomenological theory
is positioned at the real $z$ axis at $E=-r$ in the rigorous theory
is shifted to the complex plane on the distance $\propto r^{3/2}\ll r$
from the real axis, so as $r\to0$ the width of the peaked structure in
the imaginary part of GF shrinks to zero and acquires delta-function-like
shape. It will dominate the integrand at not too large values of the time
variable $t$ and the NN pair density will behave as $\simeq e^{-rt}$,
thus justifying the phenomenological approach for tightly bound I-v pairs.

\begin{figure}
\begin{center}
\includegraphics[viewport = 0 0 300 252, scale = 0.85]{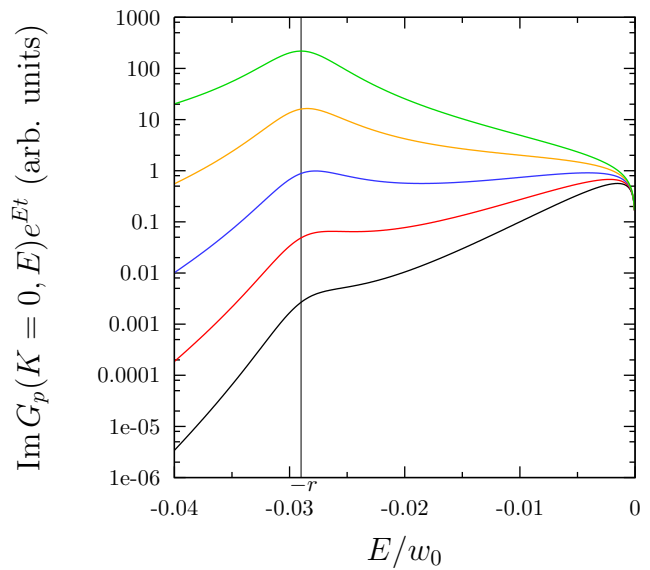}
\end{center}
\caption{\label{fig7-2}(Color online) The integrand of Eq.\
(\ref{e-rt}) for different values of the time variable for $G_p$ from Eq.\
(\ref{Gp4fm}) and $w_3=10^{-2}w_0$. Curves from top to bottom correspond
to $t=$ 10, 100, 200, 300, and 400 units of $w_0^{-1}$.}
\end{figure}

However, at small but finite values of $r$ the exponential factor
in the integrand sooner or later will suppress the quasiparticle
contribution, so at very large times the singularity at $E=0$ will still
determine the $t\to\infty$ asymptotic behavior.  This is illustrated
in Fig.\ \ref{fig7-2} where the integrand of Eq.\ (\ref{e-rt})
is shown at several stages of the evolution.

As is seen, at some stage the end-point contribution becomes dominant
and the quasiparticle behavior ceases to define the physics of the pair
dissociation.  This can be used for establishing the range of validity
of the phenomenological theory by estimating the time interval during
which the quasiparticle picture remains adequate for a given value of the
decay rate $r$. It seems reasonable to assume that the peaked structure
dominance ends when the value at the local maximum of the integrand near
$E=-r$ becomes equal to the maximum near $E=0$. The value of the latter
is found by standard means from the approximate expression
\begin{equation}
	\mbox{Im}\,G_p({\bf K=0},E^-)e^{Et}|_{E\to0}\approx	
	C_r\sqrt{|E|}e^{Et}/r.
	\label{E->0}
\end{equation}
Up to $O(1)$ factors it scales at small $r$ and large $t$ as $\sim
1/(r\sqrt{t})$.  Near the peak at $E=-r$ the integrand behaves as $\sim
e^{-rt}/r^{3/2}$ so the crossover time between the two regimes scales as
\begin{equation}
	t_c\sim -\ln r/r.
	\label{t_c}
\end{equation}
Thus, for $t<t_c$ the phenomenological description should be valid
while for larger times the power-law behavior should be observed.
In more physical terms this time interval can be estimated from the
exponential decay law in Eq.\ (\ref{e-rt}) that describes the evolution
of the density of bound pairs.  Taking the initial density as unity,
at time $t_c$ according to Eqs.\ (\ref{t_c}) and (\ref{e-rt}) it
will decrease to $r$. In Table II in I several systems with strong
I-v binding were identified on the basis of available first-principles
data.\cite{wu_high-throughput_2016,Wu2017} The parameter $w_0/w_4$ which
is proportional to $r$ when $w_4$ is large varied in those systems in
the range $\sim10^4-10^7$ which suggests that the exponential decay
law may remain valid in the pair density range in several orders of
magnitude from unity to $10^{-4}-10^{-7}$.  But there exists a natural
time limit for the pair decay to be observable. It is defined by the
association rate $g$ which defines the average association time $t_a$
Eq.\ (\ref{t_enc}). When $t_c$ is comparable to $t_a$, the new pairs in
the tight-binding case will associate before the endpoint singularity
contribution becomes detectable, so experimentally only the exponential
behavior may be seen.

In the opposite case when the decay rate is of order unity the square
root behavior in the integrand of Eq.\ (\ref{e-rt}) will also extend
over $O(1)$ region of $E$ near $E=0$ (see the lower panel in Fig.\
\ref{fig6-2}), so the power-law decay of the density will cover
practically the whole relevant time range.  Thus, in these widely
separated regions of $r$ values the density dissociation will follow
two qualitatively different laws.

Experimental observation of the decay could be helpful in experimental
and theoretical validation of the 5FM. The usefulness of the model is
hampered by the large number of parameters in its definition which cannot
be measured experimentally. In Ref.\ \onlinecite{RUEBENBAUER199180} on
the basis of EM it was argued that in cubic lattices only two of the five
frequencies can be found.  This can be seen from the phenomenological
solution Eq.\ (\ref{sigma_K}) in which the diffusion kernel is fully
characterized by only two parameters $\lambda$ and $g$. As we saw in Sec.\
\ref{mossbauer}, the encounter approximation corresponds to setting $z=0$
in the diffusion kernel while keeping its full momentum dependence.
In contrast, in the pair GF which is closely related to the diffusion
kernel (cf.\ Eqs.\ (\ref{G=rho}) and (\ref{DzK_Z})) we have set the
momentum to zero but retained $z$-dependence intact, so it may be expected
that this would provide supplementary information to that obtainable
from the momentum dependence.  Indeed, if measured, the time evolution of
the density of the tightly bound pairs would make possible determination
of the decay rate $r$ which cannot be obtained from the parameters $g$
and $\lambda$. Specifically, in Ref.\ \onlinecite{RUEBENBAUER199180}
it has been shown that $w_2$ and $w_3$ cannot be uniquely defined from
the experimental data because they are linearly correlated and the
same data can be described by different frequency sets.  The authors
discussed Fe\underline{Cu}\cite{Fe57Cu} system with a weak I-v attraction
which is difficult to analyze because of the complexity of the general
diffusion kernel.  But in the strong binding case $r$ in Eq.\ (\ref{r})
depends on $w_3$ but do not depend on $w_2$ so the knowledge of $r$
will impose a restriction on $w_3$ that is independent of $w_2$.

In Ref.\ \onlinecite{RUEBENBAUER199180} it was suggested that information
about internals of I-v pairs can be obtained from the measurement of the
impurity hyperfine interactions.  Another prospective approach for the
experimental study of the NN pairs may provide the positron annihilation
technique (see Ref.\ \onlinecite{positrons} and references therein). In
most cases the electron density inside a vacancy depends on whether the
vacancy is associated with an impurity or is surrounded only by the host
atoms. As a consequence, the lifetime and/or the momentum distribution of
the gamma radiation produced by the annihilation of the positron caught
within the vacancy will depend on whether the vacancy is associated
or not. The contribution of the associated vacancies can be separated
from the annihilation spectra and its time evolution may provide direct
information about the decay of NN pairs.
\section{\label{conclusion}Conclusion}
In the present paper a rigorous solution of the 5FM for the FCC lattice
has been obtained. Formally the solution is exact only to the first
order in the vacancy concentration but because of the small vacancy
concentration in solids this accuracy most probably exceeds the accuracy
of the model itself; besides, in order to account for higher order
corrections that may be necessary in some cases the 5FM should first be
extended to include additional many-body interactions that contribute
to higher orders of the expansion.\cite{voglAlFe} So from a physical
standpoint the first-order solution is the best one that can be obtained
within the framework of the conventional 5FM.\cite{lidiard,LECLAIRE1978}

The model solvability is a consequence of the fact that 5FM describes
interactions between only two particles: the impurity and the vacancy
and two-body problems can usually be solved exactly. The method of
solution developed in Secs.\ \ref{exact} and \ref{solution} is quite
general and can be straightforwardly generalized on other lattices
and on more sophisticated models like those suggested in literature
for more realistic description of the diffusion.  The generalizations
proposed on the basis of physical arguments and the first principles
calculations\cite{LECLAIRE1978,wu_high-throughput_2016,Bocquet,NNN_jumps}
usually include the vacancy jumps to higher CSs and longer-ranged
I-v interactions. However, in Ref.\ \onlinecite{RUEBENBAUER199180} it
was pointed out that this would necessitate introduction of additional
phenomenological parameters while even the parameters of the conventional
5FM cannot be fully determined from experimental data.  The results
of the present paper may somewhat alleviate this difficulty. The high
accuracy of the solution should make the fit of the additional parameters
more reliable than within less rigorous approaches. Besides, in Sec.\
\ref{diss} it has been shown that the time dependence of dissociation
of the NN I-v pair density can provide an additional constraint on
the parameters so it would be desirable to develop techniques for its
experimental measurement.  It has been suggested that in addition to the
hyperfine interactions technique suggested in literature the positron
annihilation is another prospective method for the study of associated
I-v pairs. Besides, such experiments will make possible investigation
of the tight binding in the cases when the binding energy is large
but for some reason the mean pair diffusion distance $\lambda$ is
small. Fe\underline{Al} system is one such case where $\lambda$ is small
because of the small value of the I-v exchange frequency.\cite{voglAlFe}
A number of similar systems can be predicted on the basis of the
first-principles calculations\cite{wu_high-throughput_2016,Wu2017} as,
e.\ g., S\underline{Al} where $E_b=0.46$~eV but $\lambda$ is only about
1~nm at the room temperature.

The method of solution developed in the present paper is applicable only
to the models with finite-range interactions. In this case the infinite
set of REs can be reduced to a finite system of linear equations and
solved by the Cramer's rule. But because the equations are linear, a
rigorous solution in the case of long-range interactions that appear,
e.\ g., in some problems of dopant diffusion in semiconductors\cite{rmp}
should be also feasible.

A major goal of the present study was to rigorously justify the
phenomenological approach developed in I on the basis of the mobile
state concept of Ref.\ \onlinecite{cowern1990} that was introduced for
the description of dopant diffusion in silicon. By assuming that the
mobile state in the 5FM coincides with the tightly bound I-v pairs,
a complete agreement between the phenomenological expression for the
impurity GF in I and the tight-binding limit of the rigorous solution
has been established.  Thus, the possibility of observation of the most
interesting phenomena such as the NGDPs\cite{cowern1990,In/cu(001)}
and the non-Fickian diffusion\cite{cowern1990} in the FCC systems with
the vacancy mediated diffusion has been rigorously substantiated.

But the rigorous solution can be also used when the conditions of validity
of the phenomenological approach are not fulfilled. The approach is
applicable only to the macroscopic diffusion at large spatiotemporal
scale and only in the case of strong I-v attraction. These conditions
are not satisfied, for example, in such techniques as the quasielastic
neutron scattering, the coherent X-rays, and the M\"ossbauer spectroscopy
where experimental data on the impurity diffusion are gathered at
large Fourier momenta that in the real space correspond to microscopic
distances.\cite{vogl_qens_1996,leitner_atomic_2009,voglAlFe} As has
been shown in Sec.\ \ref{mossbauer}, the diffusional broadening
of the M\"ossbauer resonance in Fe\underline{Al} obtained in
the framework of the rigorous approach accords well both with
experimental data and with theoretical calculations within the
EM.\cite{voglAlFe,encounter_model0,bender,vogl_qens_1996} Besides,
it has been possible to establish quantitative relations between the
parameters of the phenomenological theory and of the EM which may be
helpful in cross-checking the data obtained within the two approaches.

Farther, in view of the ongoing advancement of experimental techniques
toward the microscopic scale, an important advantage of the rigorous solution
is that it describes the impurity density at the lattice sites, i.\ e.,
at the atomic level.  Thus, in addition to the macroscopic diffusion
describable also within the phenomenological theory the rigorous approach
is able to describe the microscopic diffusion. From the results of
Refs.\ \onlinecite{Brummelhuis1988,toroczkai1997,Pd/Cu,Co/cu(001)} it
may be concluded that the microscopic NGDPs should exist in practically
all host-impurity systems, including those with I-v repulsion. So the
phenomenon should be ubiquitous and the rigorous approach provides
adequate means for its theoretical description for any values of the
5FM parameters.

But there exist important differences between the behavior of the
NN pairs in the presence of strong binding and its absence. In
Sec.\ \ref{general_case} it has been shown that the NN pairs
decay follows the exponential law in the tight-binding case and
the power law otherwise. More important difference concerns
the non-Fickian diffusion predicted in the phenomenological
theory.\cite{cowern1990,cowern1991,cowern2003,mirabella_mechanisms_2013,I}
It originates from the fact that in bound I-v pair the vacancy mediating
the diffusion is permanently available, so the diffusion flux due to
the bound pairs is unrelated to the distribution of the surrounding
impurities and can be non-zero in their absence (see, e.\ g., the
discussion in I) or even be directed along the concentration gradient
(uphill diffusion\cite{cowern2003}).  Also, the diffusion equation for
the pair diffusion does not agree with the second Fick's law because the
pairs are unstable and this their flux (hence, the flux of the impurities
they contain) is not conserved.\cite{In/cu(001),I} But this self-sustained
diffusion can persist only during the pair lifetime and so is restricted
to distances of order $\lambda$. Fick's laws describe impurity diffusion
at the macroscopic scale, so they are hardly applicable to the microscopic
diffusion where $\lambda$ is of order of the lattice constant or 
smaller. So the non-Fickian diffusion should be sought in the systems
with tightly bound pairs and macroscopic values of $\lambda$.

In I it was shown that large $\lambda$ at the room temperature
may be found in about 20\% of impurities in the aluminum
host studied in the first-principles calculations in Refs.\
\onlinecite{wu_high-throughput_2016,Wu2017}.  Unfortunately, at present
the 5FM parameters obtained in such calculations are not reliable
enough to predict even the sign of the I-v interaction in some cases.
For example, in Fe\underline{Al} system the calculations predict
I-v repulsion\cite{Wu2017} while experimentally the interaction
is attractive and quite strong.\cite{voglAlFe} Nevertheless,
it is believed that while not reliable in concrete cases, the
first-principles calculations can describe general trends in large
classes of systems.\cite{wu_high-throughput_2016} From this standpoint
it can be expected on the basis of the aluminum database\cite{Wu2017}
that I-v attraction should exist in about half of the systems, though
in many cases rather weak. But in stochastic dynamics the strength
of the interaction is measured with respect to the temperature, so
it may be hoped that at sufficiently low temperatures there exist
FCC impurity-host systems with the vacancy-mediated diffusion
that exhibit the phenomena characteristic of the non-Fickian
diffusion similar to those observed in the impurity diffusion in
semiconductors.\cite{cowern1990,cowern1991,cowern2003,mirabella_mechanisms_2013}
\begin{acknowledgments}
I would like to express my gratitude to Hugues Dreyss\'e for support.
\end{acknowledgments}
\appendix
\section{\label{master_eq}Master equation}
The master equation (ME) \cite{van_kampen}
\begin{equation}
	\frac{dp_M(t)}{dt}=\sum_{M^\prime}\left\{W_{MM^\prime}p_{M^\prime}(t)
	-W_{M^\prime M}p_M(t)\right\}
	\label{ME}
\end{equation} 
describes time evolution of the set (or vector) of probabilities
$\{p_M(t)\}$, where $M$ denotes possible states of the system, for a
stochastic system to be found in state $M$ at time $t$.  Eq.\ (\ref{ME}
is a set of liner equations with a gain-loss structure in which
the probability for the system to be found in state $M$ grows due to
transitions from other states $M^\prime$ with the rates $W_{MM^\prime}$
(the first term on the r.h.s.)  and diminishes when the system leaves
state $M$ for some other state given by the negative second term.
More compactly ME can be written as
\begin{equation}
\frac{d{p}_M(t)}{dt}=\sum_L\tilde{W}_{ML}{p}_L(t),
\label{vector}
\end{equation}
where
\begin{equation}
\tilde{W}_{ML}=W_{M\not=L}-\bigg(\sum_{M^\prime\not=M}
W_{M^\prime M}\bigg)\delta_{ML}.
\label{barW}
\end{equation}
The inequality signs here can be dropped if we agree that $W_{MM}=0$ for
all $M$. This is legitimate because Eq.\ (\ref{ME}) does not depend on the
diagonal elements of $W$. From Eq.\ (\ref{barW}) it is easy to see that
\begin{equation}
\sum_M\tilde{W}_{ML}=0
\label{sum=0}
\end{equation}
which means that the total probability $\sum_Mp_M$ is conserved 
by ME, as follows from Eq.\ (\ref{vector}).

The ME is an efficient tool for studying systems with a small number of
states, like the I-v pair that consists of one impurity and one vacancy
that was treated with the use of ME in Sec.\ 3 in I. In many-body
systems, however, the number of variables, such as particle positions,
for example, is very large and both the probability distribution $p_M$
and the ME become too cumbersome to deal with because they would account
for all fluctuations in the system though in practice one is usually
interested only in a few average quantities.\cite{van_kampen} In such
cases a reduced description by means of the rate equations (REs) for
the mean values of interest is usually more appropriate.

Let us derive a RE for the ensemble average of some fluctuating quantity $A$
\begin{equation}
	\langle A\rangle = \sum_MA_Mp_M(t).
	\label{average}
\end{equation} 
For the purposes of the present paper it would be sufficient to consider
$A$ that does not explicitly depend on time. The rate of change of the
average is obtained by multiplying both sides of Eq.\ (\ref{vector})
by $A_M$ and by summing over $M$ as
\begin{equation}
	\frac{d\langle A\rangle}{dt}=\langle B\rangle,
	\label{dA/dt}
\end{equation}
where
\begin{equation}
	B_M=\sum_LA_L\tilde{W}_{LM}.
	\label{B_M}
\end{equation}
As is seen, new average value appeared on the r.h.s.\ to which an
additional RE can be derived and by recursion an infinite system of REs
will ensue in a general case.  So  to obtain a finite closed system of
REs some approximations would be necessary.  In the main text of the
article this was achieved by restricting consideration by the linear
terms in the vacancy concentration.  In Appendix \ref{free-vacancy}
below a concrete example of application of the ME will be presented.
\section{\label{free-vacancy}Vacancy diffusion in perfect crystal}
The problem of a vacancy diffusing in a periodic lattice is a simple
example of application of the ME. Let us denote the probability for
the vacancy to be found on site ${\bf n}$ at time $t$ as $P_{\bf n}(t)$
and assume that initially the vacancy occupies site ${\bf 0}$
\begin{equation}
	P_{\bf n}(t=0) = \delta_{\bf n0}.
	\label{P(0)}
\end{equation}
The vacancy jumps between NN sites with the same rate $w_0$ so the matrix
in Eq.\ (\ref{barW}) is
\begin{equation}
	\tilde{W}_{\bf n m}^0 = w_0\bigg(\sum_{\bf e} 
	\delta_{\bf n+e, m}-12\delta_{\bf n, m}\bigg),
	\label{W_vacancy}
\end{equation}
where ${\bf e}$ is the set of 12 vectors that connect NN sites on the FCC 
lattice.  Substituting Eq.\ (\ref{W_vacancy}) into Eq.\ (\ref{ME}) one gets
\begin{eqnarray}
	\label{dP/dt}
	\frac{dP_{{\bf n}}(t)}{dt}&=&\sum_{\bf m}\tilde{W}_{\bf n m}^0
	P_{{\bf m}}(t) \nonumber\\
	&=&w_0\left(\sum_{\bf e}P_{{\bf n+e}}(t)
	-12P_{{\bf n}}(t)\right).
\end{eqnarray}
Because the equation has constant coefficients, it can be solved with
the use of the LF transform Eq.\ (\ref{LFdef}).  Applying it to Eq.\
(\ref{dP/dt}) and solving the resulting algebraic equation one arrives
at the solution
\begin{equation}
	P({\bf K},z)=\frac{1}{z+w_0a^2\epsilon_{ \bf K}},
	\label{PKz}
\end{equation}
where 
\begin{equation}
\epsilon_{\bf K} = a^{-2}\sum_{{\bf e}}\left[1-\exp({ia{\bf 
K}\cdot{\bf e}}/2)\right]
	\label{epsilon}
\end{equation}
is defined in such a way that at small $|{\bf K}|$ $\epsilon_{\bf K}\simeq
{\bf K}^2$, so that the coefficient before $\epsilon_{\bf K}$ in Eq.\
(\ref{PKz}) can be identified with the vacancy diffusion constant
\begin{equation}
	D_v=w_0a^2.
	\label{D_v}
\end{equation}

In Sec.\ \ref{solution} we need expressions for the inverse Fourier
transform of $P({\bf K},z)$
\begin{equation} 
\label{P}
P_{\bf n}(z) = \frac{1}{N}\sum_{ \bf K}
\frac{\exp[-ia{ \bf K}\cdot{\bf n}/2]}
{z+a^2w_0\epsilon_{ \bf K}}.
\end{equation} 
which are the well-known lattice GF and Watson's integrals.  In Refs.\
\onlinecite{GreenFCC,morita1975,Joyce2011} sophisticated analytical
relations between $P_{\bf n}(z)$ and the complete elliptic integrals were
derived that make possible efficient techniques of their calculation
at arbitrary complex values of $z$.  For $z\geq0$ that is sufficient
in the majority of calculations in the present paper accurate values
of all necessary $P_{\bf n}(z)$ can be found by direct numerical
integration of Eq.\ (\ref{P}) with the use of the Monkhorst-Pack
method.\cite{special_points}

From Eq.\ (\ref{dP/dt}) a useful relation between $P_{\bf n}(z)$ can
be derived
\begin{equation}
	(z+12w_0)P_{\bf n}=\delta_{\bf 0n}+w_0\sum_{\bf e}P_{\bf n+e}
	\label{relations}
\end{equation}
which can be used to express $P_{\bf n}$ with a large value of $|{\bf n}|$
through $P_{\bf m}$  with $|m|<|n|$.\cite{morita1975}  Efficiency of this
technique is enhanced by the fact that $P_{\bf n}$ satisfy all symmetries
of the cubic point group so that all $P_{\bf n}$ with any permutations
of the components of vector ${\bf n}$ with any signs are equal.
\section{\label{dyson_eq} Dyson's equation}
From the standpoint of the Mori-Zwanzig memory-function approach the
impurity GF studied in the main text is, up to normalization, the
impurity autocorrelation function and so can be treated within this
formalism.\cite{forster} The Dyson equation for the autocorrelation
function that we are interested in is derived as follows.  First one
needs to re-write ME Eq.\ (\ref{ME}) in the vector-matrix notation as
\begin{equation}
\frac{d}{dt}|t\rangle =\tilde{W}|t\rangle
\label{schroedinger}
\end{equation}
where vector $|t\rangle$ has as its components the probabilities $p_M(t)$
and $\tilde{W}$ is now the evolution operator acting on the vector.
The probability normalization $\sum_Mp_M(t)=1$ in the vector notation
can be expressed as the scalar product
\begin{equation}
\langle1|t\rangle=1,
\label{1xt}
\end{equation}
where the components of vector $\langle1|$ are equal to unity for all
$M$, so the product is just the sum over all states.
Eq.\ (\ref{schroedinger}) can be formally solved as
\begin{equation}
|t\rangle=e^{\tilde{W}t}|t=0\rangle,
\label{t=0}
\end{equation}
where we assumed that $|t=0\rangle$ is the initial state of the system 
and operator $\tilde{W}$ is time-independent.

To proceed farther we introduce two equal size sets of the ``bra'' 
\begin{equation}
\langle\bar{b}_{\bf l}|=\langle1|\hat{\imath}_{\bf l}
\label{barBm}
\end{equation}
and of the ``ket'' vectors 
\begin{equation}
|b_{\bf m}\rangle=\hat{\imath}_{\bf m}|0\rangle,
\label{Bm}
\end{equation}
where $\hat{\imath}_{\bf n}=0,1$ is the impurity occupation number on site 
${\bf n}$. The scalar product of the vectors 
\begin{equation}
\langle\bar{b}_{\bf l}|b_{\bf m}\rangle=\langle1|\hat{\imath}_{\bf l}
\hat{\imath}_{\bf m}|0\rangle=N^{-1}\delta_{\bf lm}
\label{scalar-product}
\end{equation}
is found as follows. The Kronecker symbol appears because there is only
one impurity in the system, so different sites cannot be
occupied simultaneously; $\langle1|\hat{\imath}_{\bf l}|0\rangle$
is the probability of finding the impurity at site ${\bf l}$ which is
equal to $1/N$ because all sites are equivalent.

Following the standard procedure\cite{forster} one introduces
the projection operators ${\cal P}$ and ${\cal Q}$ defined as
\begin{equation}
{\cal P}=N\sum_{\bf n}|{b}_{\bf n}\rangle\langle \bar{b}_{\bf n}|=1-{\cal Q},
\label{projectors}
\end{equation}
which satisfy the usual conditions ${\cal P}^2={\cal P}$, etc. Besides, as 
is easy to see,
\begin{equation}
{\cal Q}|{b}_{\bf n}\rangle=0
\label{Qb=0}
\end{equation}
for any ${\bf n}$. 

The correlation function
\begin{equation}
C_{\bf lm}(t)=\langle\bar{b}_{\bf l}|e^{\tilde{W}t}|{b}_{\bf m}\rangle
\label{Gij(t)}
\end{equation}
describes the probability of finding the impurity at site ${\bf m}$
in the initial state and at site ${\bf l}$ at later time $t$. As
is seen, this definition differs from the definition of GF in only one
point: in the probability of finding the impurity in the initial state
which is equal to $1/N$. In the definition of GF this probability is
defined as the certainty, i.\ e., the impurity is at site ${\bf m}$
with probability one. Thus,
\begin{equation}
G_{\bf lm}(t)=NC_{\bf lm}(t).
\label{G=C}
\end{equation}

The Laplace transforms of the correlation function Eq.\ (\ref{Gij(t)})
can be formally calculated as
\begin{equation}
C_{\bf lm}(z)=\int_0^\infty dt e^{-zt}C_{\bf lm}(t)=
\langle\bar{b}_{\bf l}|\frac{1}{z-\tilde{W}}|{b}_{\bf m}\rangle.
\label{Gij(z)}
\end{equation}
Now with the use of Eq.\ (\ref{Qb=0}) and the operator identities 
$1={\cal P+Q}$ and
\begin{equation}
\frac{1}{x-y}=\frac{1}{x}+\frac{1}{x}y\frac{1}{x-y}
\label{x-y-1}
\end{equation} 
Eq.\ (\ref{Gij(z)}) can be transformed as\cite{forster} 
\begin{equation}
C_{\bf lm}(z)=\langle\bar{b}_{\bf l}|\left(\frac{1}{z}
+\frac{1}{z-\tilde{W}{\cal Q}}
\tilde{W}{\cal P}\frac{1}{z-\tilde{W}}\right)|{b}_{\bf m}\rangle.
\label{G(z)2}
\end{equation}
Using Eq.\ (\ref{projectors}) one finds
\begin{equation}
{\cal P}\frac{1}{z-\tilde{W}}|{b}_{\bf m}\rangle 
=N\sum_{\bf n}|{b}_{\bf n}\rangle\langle 
\bar{b}_{\bf n}|\frac{1}{z-\tilde{W}}|{b}_{\bf m}\rangle
\label{PC}
\end{equation}
Substituting this into Eq.\ (\ref{G(z)2}) and using Eqs.\
(\ref{scalar-product}), (\ref{G=C}), and (\ref{Gij(z)}) one gets
\begin{equation}
G_{\bf lm}(z)=\frac{\delta_{\bf lm}}{z}+N\langle\bar{b}_{\bf l}|
\frac{1}{z-\tilde{W}{\cal Q}}
\tilde{W}|{b}_{\bf n}\rangle G_{\bf nm}(z).
\label{dyson2}
\end{equation}
This is an alternative form of the Dyson equation in lattice coordinates. To 
see this we first apply the identity Eq.\ (\ref{x-y-1}) to Eq.\ (\ref{dyson})
to get
\begin{equation}
G({\bf K},z)=\frac{1}{z-\Sigma({\bf K},z)}=\frac{1}{z}+\frac{1}{z} 
\Sigma({\bf K},z)G({\bf K},z).
\label{dyson3}
\end{equation}
This would coincide with the Fourier transformed Eq.\ (\ref{dyson2}) provided
the self-energy is 
\begin{equation}
\Sigma({\bf K},z)=\sum_{\bf l}e^{-i{\bar{\bf K}\cdot {\bf l}}}
N\langle\bar{b}_{\bf l}|
\frac{z}{z-\tilde{W}{\cal Q}}
\tilde{W}|{b}_{\bf 0}\rangle
\label{dyson4}
\end{equation}
where the choice ${\bf n=0}$ has been made to simplify  the expression
and $\bar{\bf K}=a{\bf K}/2$. Eq.\ (\ref{dyson4}) is rather complicated
because it is exact to all orders in $c_v$. But as was argued in the
main text, only the first order term is physically sound and it can be
obtained by simpler means.
\section{\label{parameters} Expressions from the phenomenological theory}
Below are listed some formulas that are needed for comparison with
the phenomenological approach in the main text.  Their derivation can
be found in I.  

In the phenomenological theory the strong I-v attraction is characterized
by large binding energy $E_b>0$ which defines the equilibrium density
of the vacancies on the impurity NN sites as
\begin{equation}
c_{NN}^{(eq)}\simeq c_ve^{E_b/k_BT}
	\label{c^eq}
\end{equation}
and enters the detailed balance condition that imposes the 
restriction on the jump frequencies\cite{LECLAIRE1978,manning,philibert}
\begin{equation}
	\frac{w_3}{w_4}\simeq\exp\biggl(-\frac{E_b}{k_BT}\biggr)\to0.
	\label{w3/w4}
\end{equation}

In the phenomenological theory the diffusion kernel at $z=0$ is
\begin{equation}
	\Sigma^{(ph)}({\bf K},0)\approx
-g\left(1-\frac{1}{1+(\lambda{\bf K})^2}\right)
	\label{sigma_K}
\end{equation}
(see Eq.\ (69) in I). The mean pair diffusion distance is 
\begin{equation}
	\lambda=\sqrt{D_m/r}
	\label{lambda}
\end{equation}
(Eq.\ (27) in I), where the pair diffusion constant
\begin{equation}
	D_m=\frac{w_2(w_1+w_3)a^2}{12(w_1+w_2+w_3)}
=\frac{w_2}{12}f_\infty a^2
	\label{Dm}
\end{equation}
and the pair decay rate
\begin{equation}
	r=7w_3p_\infty(w_4/w_0)
	\label{r}
\end{equation}
were introduced in I in Eqs.\ (24) and (28); 
\begin{equation}
	f_\infty=\frac{w_1+w_3}{w_1+w_2+w_3}
	\label{f_infty}
\end{equation}
is the correlation factor $f$ in the strong coupling limit (Eq.\ (7)
in I); $p_\infty$ is the probability of the vacancy in the I-v pair to
diffuse from the impurity NN site at the spatial infinity.
\section{\label{4FM}The four frequency model (4FM)}
In the calculations of the diffusion constant for the 5FM it has been
noted\cite{koiwa1983,Bocquet}  that the matrices needed in the
derivation greatly simplify when $w_4=w_0$. It turns out that in the
general case of the diffusion kernel the matrices in the determinants
in the Cramer's solution Eqs.\ (\ref{cramers}), (\ref{det0}), and
(\ref{DzK_Z}) also become much simpler in this case.  When $w_4=w_0$
the nonzero matrix elements remain only in the first three columns of
matrix $U$ in Eq.\ (\ref{U}) and as a consequence in the matrix product
$HU$. The physical reason for this is clear. As can be seen from Fig.\
\ref{fig1-2}, the region of influence of the impurity shrinks in
this case to only the first CS which in the axisymmetric case contains
only three components of vector $\vec{\rho}$ on the r.h.s.\ of Eq.\
(\ref{rhs}), so the set of equations, hence, the size of determinants
in Eq.\ (\ref{cramers}) can be reduced to three:
\begin{widetext}
\begin{eqnarray}
\label{det}
&&\Delta|_{w_4=w_0}\equiv\left|\begin{array}{ccc}
d_{11}&d_{12}&d_{13}\\
d_{21}&d_{22}&d_{23}\\
d_{31}&d_{32}&d_{33}
\end{array}\right|= \\
&&\left|\begin{array}{c|c|c}
1+4H_{10}-H_{11}+2(H_{11}-H_{12})\bar{w}_1+(H_{11}-H_{13}e^{-i\bar{K}})w_2
+A\bar{w}_3& d_{12}=d_{21}(\bar{K}=0)&d_{31}^{(-K)}\\\hline
4H_{10}-H_{21}+2(H_{21}-H_{22})\bar{w}_1+(H_{21}-H_{23}e^{-i\bar{K}})w_2
+B\bar{w}_3&
1+4H_{10}-H_{22}+4(H_{22}-H_{21})\bar{w}_1+E\bar{w}_3&d_{21}^{(-K)}\\\hline
4H_{10}-H_{13}+2(H_{13}-H_{23})\bar{w}_1+(H_{13}-H_{11}e^{-i\bar{K}})w_2
+C\bar{w}_3& d_{12}&d_{11}^{(-K)}
\end{array}\right|,\nonumber
\end{eqnarray}
\end{widetext}
where the vertical lines on both sides of a matrix denote its determinant
and the superscripts designate the change of sign ${K}\to-{K}$.

Additional simplification takes place in determinant $\Delta_1$ in the
numerator of the Cramer's solution Eq.\ (\ref{cramers}) due to the fact
that in $3\times3$ determinant only three components of the r.h.s.\ vector
Eq.\ (\ref{Prho}) corresponding to NN sites are needed. But all these
components are equal because they depend on the same function $P_{\bf
e}=P_{011}$ so that similar to Eq.\ (\ref{lim_z}) the determinant can
be calculated with the first column in $\Delta$ replaced by unities
\begin{equation}
\bar{\Delta}_1=\Delta|_{d_{i1}=1,i=1-3}
\label{delta_1}
\end{equation}
and then multiplied by the factor
\begin{equation}
\eta_{NN}=(\tilde{P}\vec{c}_{0})_{\bf e}=\frac{c_v}{z}+[(c_{NN}-c_v) 
(12+\frac{z}{w_0})-c_v]P_{011}.
	\label{DPrho_1}
\end{equation}
Further details of the calculation of determinants $\Delta$ and 
$\bar{\Delta}_1$ are given in Sec.\ \ref{dets} below.
Substituting Eqs.\ (\ref{det_S1}) and (\ref{DET4fm}) into Eqs.\
(\ref{cramers}) and (\ref{zG}) one gets
\begin{widetext}
\begin{equation}
G(K,z)=\frac{1}{z}\left(1-\frac{8w_2\eta_{NN}A_1A_2(1-\cos  \bar{K})}
{A_1(A_2+2C_2w_2)A_3 +2w_2\big\{A_3[C_2(A_4+2H_{12}w_1)+C_4w_3] 
	-A_2C_3A_4\big\}(1-\cos\bar{K})}\right)
\label{G4fm(K,z)}
\end{equation}
\end{widetext}
(notation is explained in Sec.\ \ref{dets}).  This expression is exact
up to the first orders in $c_v$ and can be used to confirm by explicit
formulas the qualitative discussion of Sec.\ \ref{general_case} and the
numerical findings of Sec.\ \ref{D-r-Dm}, as well as relevant expressions
obtained in I within the phenomenological theory. However, because in
the 4FM $w_4=w_0$, only the small-$w_3$ case of the tight binding can
be considered.

To establish correspondence with the phenomenological theory we have
to find the behavior of GF in Eq.\ (\ref{G4fm(K,z)}) in the region
of validity of the phenomenological approach which is restricted
to small values of $z$, ${\bf K}^2$, and $w_3$.  To this end below
we will approximate all necessary expressions by only leading terms
in the small quantities.  It is to be noted that we do not intend to
develop a systematic expansion but simply keep the most important terms
to simplify Eq.\ (\ref{G4fm(K,z)}) to reveal its physical content. In
case the accuracy might seem to be insufficient, additional terms from
the exact expression can be added to improve the approximation.

Thus, below we will see that the terms in the numerator of the
phenomenological GF in Eq.\ (52) in I proportional to $c_{NN}$ and $g$
originate from the two leading terms in powers of $z$ in the initial
condition Eq.\ (\ref{Prho})
\begin{equation}
	\eta_{NN}\approx c_v/z+c_{NN}12P_{\bf e}=c_v/z+c_{NN}C_3|_{z=0},
	\label{DPrho-approx}
\end{equation}
where use has been made of Eq.\ (\ref{C_3}).  The approximate
expressions for $A_i$ in Eqs.\ (\ref{Ak-Ck}), (\ref{A3}), (\ref{C_3}),
and (\ref{A4-C4}) read
\begin{eqnarray}
\label{ACapprox}
A_1&\approx& 6C_1w_1\nonumber\\
A_2&\approx& 2C_2w_1\nonumber\\
A_3&\approx& C_3z+w_3\nonumber\\
A_4&\approx& -2C_1w_1.
\end{eqnarray}
As is seen, $A_3$ to the leading order is linear in both $z$ and $w_3$ so
the first term in the braces in the denominator of Eq.\ (\ref{G4fm(K,z)})
can be dropped because it is further multiplied by $K^2$ and thus
will give a quadratic contribution. Under the approximations Eq.\
(\ref{G4fm(K,z)}) simplifies to
\begin{equation}
	G(K,z)=\frac{1}{z}\left(1-\frac{4w_2\eta_{NN}f_{w_3=0}\bar{K}^2}
	{A_3-w_2f_{w_3=0}C_3(A_4/A_1)\bar{K}^2}\right),
	\label{G4fm1}
\end{equation}
where 
\begin{equation}
	f|_{w_3=0}=\frac{A_2}{A_2+2C_2w_2}=\frac{w_1}{w_1+w_2}.
	\label{fw30}
\end{equation}
Now substituting the approximate expressions Eqs.\ (\ref{ACapprox}) into
Eq.\ (\ref{G4fm1}) and remembering that $\bar{K}=aK/2$ one finally gets
\begin{equation}
	G(K,z)\approx\frac{1}{z}-\frac{(12\eta_{NN}/C_3)D_m^{w_3=0}K^2}{z(z+w_3/C_3+D_m^{w_3=0}K^2)},
\label{Gapprox}
\end{equation}
where $D_m$ is given by Eq.\ (\ref{Dm}) and $\eta_{NN}$ by Eq.\ 
(\ref{DPrho-approx}).

Eq.\ (\ref{Gapprox}) will agree with Eq.\ (52) from I provided that, first,
\begin{equation}
w_3/C_3|_{z\to0}=r=7w_3p_\infty(w_4/w_0=1).
\label{r=r}
\end{equation}
This is indeed the case because of the following chain of equalities 
\begin{equation}
C_3^{-1}|_{z=0}=(12P_{011})^{-1}|_{z=0}=2.90=7p_\infty(1),
	\label{C3=P=p}
\end{equation}
where the last equality was obtained in MC simulations in I.
The contributions proportional to $g$ and $c_{NN}$ in Eq.\ (52) in I
also fully agree with the 4FM expression Eq.\ (\ref{Gapprox}), as can be
shown with the use of Eqs.\ (\ref{DPrho-approx}), (\ref{C3=P=p}) and Eq.\
(57) from I
\begin{equation}
	g=84c_vw_4p_\infty,
	\label{g}
\end{equation}
where in the 4FM $w_4$ should be replaced by $w_0$.

Thus, in the case of 4FM the phenomenological limit of the impurity GF
can be rigorously obtained by analytic means.
\subsection{\label{dets}Calculation of determinants}
The coefficients of $\bar{w}_3$ in Eq.\ (\ref{det}) are the following
combinations of the matrix elements $H_{ij}$:
\begin{eqnarray}
	\label{ABCE}
&&A=7H_{11}-4H_{14}-H_{15}-2H_{17}-H_{18}-H_{1,11}\nonumber\\
&&B=7H_{21}-2H_{15}-H_{18}-H_{19}-H_{1,12}\nonumber\\ 
&&C=7H_{31}-H_{15}-4H_{16}-H_{19}-2H_{1,10}-H_{1,13}\nonumber\\
&&E=7H_{22}-2H_{25}-H_{28}-H_{29}-H_{2,12},
\end{eqnarray}
Though the determinant of a $3\times3$ matrix can be calculated
analytically with the use of known elementary formula, in the case
of Eq.\ (\ref{det}) the resulting expressions will be still too
awkward for analysis.  In this Appendix several tricks are suggested
to simplify the task. The main tool to use is the identity Eq.\
(\ref{relations}) satisfied by functions $P_{ijk}(z)$.  In Refs.\
\onlinecite{GreenFCC,morita1975,Joyce2011} it was shown that $P_{ijk}(z)$
with any indexes can be expressed through only three of them.  In our
calculations we used this possibility only partially by reducing $H_{ij}$
in Eqs.\ (\ref{ABCE}) with high indexes to lower ones as
\begin{eqnarray}
	\label{newABCE}
&&A=1+4H_{10}-(z+3)H_{11}+2H_{12}\nonumber\\
&&B=4H_{10}-(z+3)H_{12}+2H_{22}\nonumber\\
&&C=4H_{10}-(z+3)H_{13}+2H_{12}\nonumber\\
&&E=1+4H_{10}-(z+5)H_{22}+4H_{12}
\end{eqnarray}

Similarly, by replacing barred frequencies $\bar{w}_k$ in Eq.\ (\ref{det})
with $w_k-1$ (we remind that $w_0$ was set to be unity) it is possible
to cast the determinant in the form
\begin{widetext}
\begin{equation}
	\Delta=
\left|\begin{array}{l|c|c}
zH_{11}+2(H_{11}-H_{12})w_1+(H_{11}-H_{13}e^{-i\bar{K}})w_2+Aw_3
&d_{12}=d_{21}(K=0)&d_{31}^{(-K)}\\\hline
zH_{12}+2(H_{12}-H_{22})w_1+H_{12}(1-e^{-i\bar{K}})w_2+Bw_3&
zH_{22}+4(H_{22}-H_{12})w_1+Ew_3&d_{21}^{(-K)}\\\hline
zH_{13}+2(H_{13}-H_{12})w_1+(H_{13}-H_{11}e^{-i\bar{K}})w_2+Cw_3&
d_{12}&d_{11}^{(-K)}
\end{array}\right|,
\label{det_S}
\end{equation}
\end{widetext}
First let us consider the simpler case of the determinant
$\bar{\Delta}_1$ that is obtained from $\Delta$ by placing unities in the first
column in Eq.\ (\ref{det_S}). By subtracting the first line from the
second and the third ones a triangular matrix is obtained with determinant
\begin{equation}
	\bar{\Delta}_1=(d_{22}-d_{12})(d_{11}^{(-K)}-d_{31}^{(-K)}).
	\label{det_S1}
\end{equation}
Using explicit expressions for $d_{ij}$ from Eqs.\ (\ref{det_S})
and (\ref{newABCE}) it is easy to transform Eq.\ (\ref{det_S1}) to
\begin{equation}
\bar{\Delta}_1=A_1[A_2+C_2w_2(1+e^{i\bar{K}})],
	\label{det_S1-2}
\end{equation}
where
\begin{eqnarray}
	\label{Ak-Ck}
A_1 &=& d_{22}-d_{12}=w_3+C_1[z+6w_1-(z+7)w_3]\nonumber\\
C_1 &=& H_{22}-H_{12}\nonumber\\
A_2 &=&w_3+C_2[z+2w_1-(z+3)w_3]\nonumber\\
C_2 &=& H_{11}-H_{13}\nonumber
\end{eqnarray}

The calculation of $\Delta$ is more complicated but it can be simplified
by the observation that at $K=0$ all columns sum to the same value
\begin{equation}
A_3=C_3z+ [1-z(z+13)H_{10}]w_3 
\label{A3}
\end{equation}
where 
\begin{equation}
C_3 = (z+12)H_{10}.
	\label{C_3}
\end{equation}
This can be shown with the use of the identities of the kind
\begin{equation}
H_{11}+H_{12}+H_{13}=2H_{12}+H_{22}=C_3,
\label{identity}
\end{equation}
\begin{equation}
A+B+C=2B+E=1-z(z+13)H_{10},
\label{relation}
\end{equation}
etc., obtained with the use of Eq.\ (\ref{relations}). At finite $K$,
by adding the first and the third rows to the middle one the second row
will contain the following matrix elements:
\begin{equation}
[A_3+C_3w_2(1-e^{-i\bar{K}}),A_3,A_3+C_3w_2(1-e^{i\bar{K}})]
\label{2nd_row}
\end{equation}
With $A_3$ given by Eq.\ (\ref{A3}) it is seen that in the limit of
strong binding $w_3\to0$ and the diffusion limit $z,K^2\to0$ all terms
in the row are small being of the first order in all these quantities.
Because $\Delta$ can be calculated as the sum of these terms multiplied by 
corresponding minors,
\begin{eqnarray}
	&&\Delta=A_1(A_2+2C_2w_2)A_3 
	+2w_2\big(A_3[C_2(A_4\nonumber\\
	&&+2H_{12}w_1)+C_4w_3]-A_2C_3A_4\big)(1-\cos\bar{K}),
	\label{DET4fm}
\end{eqnarray}
where
\begin{eqnarray}
A_4 &=& zH_{12}-2C_1w_1+Bw_3\nonumber\\
C_4 &=& H_{13}-2C_2(2H_{10}+H_{12})
\label{A4-C4}
\end{eqnarray}
this shows that the general form of the determinant Eq.\ (\ref{product}) can
be rigorously justified at least in the 4FM.

The derivations above are rather cumbersome but their validity can be
checked numerically. Because all identities used are purely algebraical,
they can be verified for real values of both $z$ and $iK$ (i.\ e., for
complex $K$). As was pointed out in Sec.\ \ref{free-vacancy}, for $z\ge0$
all $P_{ijk}$ can be efficiently computed with the use of the special
points technique.\cite{special_points} All identities verified in this
way were found to be correct in most cases almost to the accuracy of
the double precision arithmetic for all values of $z$ and $K$ studied.
\bibliographystyle{apsrev4-1} 

\begin{thebibliography}{44}
\bibitem [{SM()}]{SM}%
  \BibitemOpen
  \href@noop {} {}\bibinfo {note} {See ancillary file H-P.txt for $H_{ij}$
  expressed in terms of $P_{lmn}$ in the form ready to be used in a computer
  code}\BibitemShut {NoStop}%
\end{thebibliography}
\end{document}